\documentclass[11pt, a4paper]{article}

\makeatletter%

\usepackage{ifthen}



\makeatletter%

\newcommand{\ifndef}[2]{\ifthenelse{\isundefined{#1}}{#2}{}}

\newcommand{\mydef}[2]{\def#1{#2}}

\newcommand{\nospell}[1]{#1}  %

\makeatletter
\newcommand{\myusepackage}[2][]{\@ifpackageloaded{#2}{} %
{\ifthenelse{\equal{}{#1}} {\usepackage{#2}} {\usepackage[#1]{#2}} }}
\makeatother

\myusepackage{amssymb}
\myusepackage{amsmath}  %

\myusepackage{amsthm}  %

{}
\myusepackage[T1,OT2,T2A]{fontenc}
\DeclareTextSymbolDefault{\CYRYAT}{OT2}
\DeclareTextSymbolDefault{\cyryat}{OT2}
\DeclareTextSymbolDefault{\CYRFITA}{OT2}
\DeclareTextSymbolDefault{\cyrfita}{OT2}
\DeclareTextSymbolDefault{\CYRIZH}{OT2}
\DeclareTextSymbolDefault{\cyrizh}{OT2}
\myusepackage[utf8]{inputenc}
\let\f\relax

\myusepackage{latexsym}
\myusepackage{amsfonts}
\myusepackage{units}    %
\myusepackage{psfrag}   %
\myusepackage{url}   %
\myusepackage{hyperref}  %
\myusepackage{hyphenat}  %
\myusepackage{microtype}   %
\myusepackage{marginnote}  %
{}  %
\myusepackage{enumitem}  %
{}
\myusepackage[dvips]{graphicx}
\myusepackage[usenames,dvipsnames]{color}
{}
{}  %
{} %

{}
\newcommand{\dgDefinition}{Definition}
\newcommand{\dgDefinitions}{Definitions}
\newcommand{\dgFact}{Fact}
\newcommand{\dgFacts}{Facts}
\newcommand{\dgQuestion}{Question}
\newcommand{\dgQuestions}{Questions}
\newcommand{\dgLemma}{Lemma}
\newcommand{\dgLemmas}{Lemmas}
\newcommand{\dgCorollary}{Corollary}
\newcommand{\dgCorollaries}{Corollaries}
\newcommand{\dgProposition}{Proposition}
\newcommand{\dgPropositions}{Propositions}
\newcommand{\dgClaim}{Claim}
\newcommand{\dgClaims}{Claims}
\newcommand{\dgTheorem}{Theorem}
\newcommand{\dgTheorems}{Theorems}
\newcommand{\dgProblem}{Problem}
\newcommand{\dgProblems}{Problems}
\newcommand{\dgRemark}{Remark}
\newcommand{\dgRemarks}{Remarks}
\newcommand{\dgConjecture}{Conjecture}
\newcommand{\dgConjectures}{Conjectures}
\newcommand{\dgResult}{Result}

\newcommand{\dgProofOf}{\proofname of}
{}

{}
\ifndef{\theorem}{\newtheorem{theorem}{\dgTheorem}}
{}

{}  %
{}  %
\ifndef{\lemma}{\newtheorem{lemma}{\dgLemma}}
\ifndef{\corollary}{\newtheorem{corollary}{\dgCorollary}}
\ifndef{\conjecture}{\newtheorem{conjecture}{\dgConjecture}}
\ifndef{\remark}{\theoremstyle{remark} \newtheorem{remark}{\dgRemark}}
\ifndef{\proposition}{\newtheorem{proposition}{\dgProposition}}
\ifndef{\claim}{\newtheorem{claim}{\dgClaim}}
\ifndef{\result}{\newtheorem{result}{\dgResult}}
\ifndef{\problem}{\newtheorem{problem}{\dgProblem}}
{}  %
{}  %

\newtheoremstyle{mydefinition}  %
{\topsep}{\topsep}  %
{\slshape}  %
{}  %
{\bfseries}  %
{.}  %
{ }  %
{}  %

\newtheoremstyle{mynotation}  %
{\topsep}{\topsep}  %
{}  %
{}  %
{\bfseries\slshape}  %
{.}  %
{ }  %
{}  %

\newtheoremstyle{myremark}  %
{\topsep}{\topsep}  %
{\slshape}  %
{}  %
{\bfseries\slshape}  %
{:}  %
{ }  %
{}  %

\newtheoremstyle{myexample}  %
{\topsep}{\topsep}  %
{\itshape}  %
{}  %
{\slshape}  %
{:}  %
{ }  %
{\ul{\thmname{#1}}}  %

\newtheoremstyle{myclaims}  %
{\topsep}{\topsep}  %
{\slshape}  %
{}  %
{\bfseries\itshape}  %
{.}  %
{ }  %
{\thmname{#1}\thmnumber{ \!#2}\ifthenelse{\equal{}{#3}} %
{}{\textnormal{ \!(#3)}}}  %

{} %
\ifndef{\notation}
{\theoremstyle{mynotation}\newtheorem*{notation}{Notation}}
{} %
{\theoremstyle{myremark}}
{} %
\ifndef{\definition}
{\theoremstyle{mydefinition}\newtheorem{definition}{\dgDefinition}}
\ifndef{\example}
{\theoremstyle{myexample}\newtheorem{example}{Example}}
{\theoremstyle{myclaims}

\ifndef{\fact}{\newtheorem{fact}{\dgFact}}
\ifndef{\question}{\newtheorem{question}{\dgQuestion}}
}
{} %
{} %
{} %

{}
\newtheoremstyle{anystatement}{\topsep}{\topsep}{\itshape}{}{\bfseries}{.}{ }{\anystatementname}
{\theoremstyle{anystatement}}
\newcommand{\anystatementname}{}

{}

\newcommand{\newident}[3][*]{\ifthenelse{\equal{*}{#1}}%
{\newcommand{#2}[1][]{\Ensuremath{\mathit{#3##1}}}}%
{\newcommand{#2}[1][]{\Ensuremath{\mathit{#3}}}}%
}

\newcommand{\newmat}[3][*]{\ifthenelse{\equal{*}{#1}}%
{\newcommand{#2}[1][]{\Ensuremath{#3##1}}}%
{\newcommand{#2}[1][]{\Ensuremath{#3}}}%
}

\newcommand{\providemat}[3][*]{\ifthenelse{\equal{*}{#1}}%
{\providecommand{#2}[1][]{\Ensuremath{#3##1}}}%
{\providecommand{#2}[1][]{\Ensuremath{#3}}}%
}

\newcommand{\newmatop}[2]{\mydef{#1}{\operatorname{#2}}}

\newcommand{\newfunction}[2]{%
\newcommand{#1}[2][*]{\ifthenelse{\equal{*}{##1}}%
{\Ensuremath{#2{\left(##2\right)}}}%
{#2(##2)}}%
}

\newcommand{\MyMakeTheoMacros}[3]{
\newcommand{#2}[2][]{\ifthenelse{\equal{}{##1}}
{\begin{#1} ##2 \end{#1}}
{\begin{#1}\label{##1} ##2\end{#1}}}
\newcommand{#3}[3][]{\ifthenelse{\equal{}{##1}}
{\begin{#1}[\e{##2}] ##3 \end{#1}}
{\begin{#1}[\e{##2}]\label{##1} ##3\end{#1}}}
}

\makeatletter
\newtheorem*{rep@theorem}{\rep@title}
\newcommand{\newreptheorem}[2]{%
\newenvironment{rep#1}[1]{%
\def\rep@title{#2 \ref{##1}}%
\begin{rep@theorem}}%
{\end{rep@theorem}}}
\makeatother

\newcommand{\MyMakeDupTheoMacros}[7]{
\MyMakeTheoMacros{#1}{#2}{#3}
\newreptheorem{#1}{#6}
\newcommand{#4}[3]{
\newcommand{##2}{##3}
\begin{#1}\label{##1} ##2\end{#1}}
\newcommand{#5}[4]{
\newcommand{##2}{##4}
\begin{#1}{\e{##3}}\label{##1} ##2\end{#1}}
\newcommand{#7}[2]{\begin{rep#1}{##1} ##2 \end{rep#1}}
}

{}
\newcommand{\MyMakeRefMacros}[3]{\newcommand{#1}[2][]
{\ifthenelse{\equal{}{##1}}{#2~\ref{##2}}{#3~\ref{##1} and~\ref{##2}}}}

\newcommand{\MyMakeEqRefMacros}[3]{\newcommand{#1}[2][]
{\ifthenelse{\equal{}{##1}}{#2~\eqref{##2}}{#3~\eqref{##1} and~\eqref{##2}}}}
{}

{}  %
\newcommand{\bibentry}[8]{
{}\bibitem[\nospell{#8}]{#1} {\textup #3}.{}
\ifthenelse{\equal{}{#6}}
{\newblock \textrm{#4.} \newblock {\em #5}, #7.}
{\newblock \textrm{#4.} \newblock {\em #5, #6}, #7.}
}

{} %

\MyMakeTheoMacros{fact}{\fct}{\nfct}

\MyMakeRefMacros{\fctref}{\dgFact}{\dgFacts}

\MyMakeTheoMacros{question}{\quest}{\nquest}

\MyMakeRefMacros{\questref}{\dgQuestion}{\dgQuestions}

\MyMakeTheoMacros{notation}{\nota}{\nnota}

{} %
\MyMakeDupTheoMacros{lemma}
{\lem}{\nlem}{\lemdup}{\nlemdup}{\dgLemma}{\lemrep}
{}

\MyMakeRefMacros{\lemref}{\dgLemma}{\dgLemmas}

\MyMakeDupTheoMacros{corollary}
{\crl}{\ncrl}{\crldup}{\ncrldup}{\dgCorollary}{\crlrep}

\MyMakeRefMacros{\crlref}{\dgCorollary}{\dgCorollaries}

\MyMakeTheoMacros{proposition}{\prp}{\nprp}

{}
\newtheorem*{prp*}{\e{\dgProposition}}

{}

\MyMakeRefMacros{\prpref}{\dgProposition}{\dgPropositions}

\MyMakeDupTheoMacros{my_claim}
{\clm}{\nclm}{\clmdup}{\nclmdup}{\dgClaim}{\clmrep}

\MyMakeRefMacros{\clmref}{\dgClaim}{\dgClaims}

\MyMakeDupTheoMacros{theorem}
{\theo}{\ntheo}{\theodup}{\ntheodup}{\dgTheorem}{\theorep}

\MyMakeRefMacros{\theoref}{\dgTheorem}{\dgTheorems}

\MyMakeTheoMacros{definition}{\defi}{\ndefi}

\MyMakeRefMacros{\defiref}{\dgDefinition}{\dgDefinitions}

\MyMakeTheoMacros{problem}{\prob}{\nprob}

\MyMakeRefMacros{\probref}{\dgProblem}{\dgProblems}

\MyMakeTheoMacros{remark}{\rem}{\nrem}

\MyMakeRefMacros{\remref}{\dgRemark}{\dgRemarks}

\MyMakeTheoMacros{conjecture}{\conj}{\nconj}

\MyMakeRefMacros{\conjref}{\dgConjecture}{\dgConjectures}

\renewcommand{\qedsymbol}{$\blacksquare$}

\newcommand{\prfstart}[1][]{\ifthenelse{\equal{}{#1}}%
{\begin{proof}\renewcommand{\qedsymbol}{$\blacksquare$}}%
{\begin{proof}[\dgProofOf\ #1]%
\renewcommand{\qedsymbol}{$\blacksquare_{\mbox{\it{\scriptsize{#1}}}}$}}%
}
\newcommand{\prfend}[1][*]{%
\ifthenelse{\equal{}{#1}}{\renewcommand{\qedsymbol}{$\blacksquare$}}{}%
\ifthenelse{\equal{*}{#1}}{}%
{\renewcommand{\qedsymbol}{$\blacksquare_{\mbox{\it{\scriptsize{#1}}}}$}}%
\end{proof}\renewcommand{\qedsymbol}{$\blacksquare$}%
}

\newcommand{\sect}[2][]{
\ifthenelse{\equal{*}{#2}}
{\section*}
{\ifthenelse{\equal{}{#1}}
{\section{#2}}
{\section{#2}\label{#1}}
}
}

\newcommand{\ssect}[2][]{
\ifthenelse{\equal{*}{#2}}
{\subsection*}
{\ifthenelse{\equal{}{#1}}
{\subsection{#2}}
{\subsection{#2}\label{#1}}
}
}

\newcommand{\sssect}[2][]{
\ifthenelse{\equal{*}{#2}}
{\subsubsection*}
{\ifthenelse{\equal{}{#1}}
{\subsubsection{#2}}
{\subsubsection{#2}\label{#1}}
}
}

\MyMakeRefMacros{\chref}{Chapter}{Chapters}

\MyMakeRefMacros{\sref}{Section}{Sections}

\MyMakeRefMacros{\ssref}{Subsection}{Subsections}

\MyMakeRefMacros{\sssref}{Subsection}{Subsections}

\MyMakeRefMacros{\figref}{Figure}{Figures}

\newcommand{\IfMathMode}[2]{\ifmmode{#1}\else{#2}\fi}

\newcommand{\Ensuremath}{\ensuremath}

\newcommand{\fbr}[1]{\IfMathMode%
{#1}{$#1$}}                     %

\newcommand{\fnbr}[1]{\mbox{\fbr{#1}}}  %

\newcommand{\fla}[2][*]{\ifthenelse{\equal{}{#1}}{\fbr{#2}}{\fnbr{#2}}}
\newcommand{\f}{\fla}

\newcommand{\bfla}[2][]{\mbox{\fbr{\mathbf{#2}}}}
\newcommand{\fb}{\bfla}

\newcommand{\malabel}[1]{\addtocounter{equation}{1}\tag{\theequation}\label{#1}}
\newcommand{\mal}[2][]{%
\ifthenelse{\equal{}{#1}}%
{\begin{align*} #2 \end{align*}}%
{\ifthenelse{\equal{P}{#1}}%
{\begingroup\allowdisplaybreaks\begin{align*} #2%
\end{align*}\endgroup}%
{\begin{align*} \malabel{#1} #2 \end{align*}}%
}%
}

\newcommand{\m}{\mal}

\newcommand{\mac}{\substack}

\MyMakeEqRefMacros{\equref}{Equation}{Equations}

\MyMakeEqRefMacros{\expref}{Expression}{Expressions}

\MyMakeEqRefMacros{\inequref}{Inequality}{Inequalities}

\newcommand{\bref}[1]{(\ref{#1})}

\newcommand{\twocase}[4]%
{\begin{cases} #1 &\txt{#2}\\ #3 &\txt{#4}\end{cases}}

\newcommand{\thrcase}[6]%
{\begin{cases} #1 &\txt{#2}\\ #3 &\txt{#4}\\ #5 &\txt{#6}\end{cases}}

\providecommand{\middle}{\big}

\newcommand{\chs}{\genfrac(){0cm}{}}  %
\newmatop{\supp}{supp}   %
\providecommand{\cupdot}{\mathbin{\cdot\mkern-8.45mu\cup}}

\newcommand{\h}[2][]{\ifthenelse{\equal{}{#2}}%
{\mathop H_{#1}}%
{\mathop H_{#1}{\l({#2}\r)}}}
\newcommand{\hh}[3][]{\mathop H_{#1}%
{\l({#2}\vphantom{|_1^1}\md|\vphantom{|_1^1}{#3}\r)}}

\newcommand{\hbin}[2][]{h_2#1\l(#2\r)}        %

\newcommand{\I}[3][]{\ifthenelse{\equal{}{#1}}%
{\mathbf{I}{\left[{#2}\vphantom{|_1^1}:\vphantom{|_1^1}{#3}\right]}}%
{\mathop{\mathbf{I}}_{#1}{\left[{#2}\vphantom{|_1^1}:\vphantom{|_1^1}{#3}\right]}}%
}
\newcommand{\Ii}[4][]{\ifthenelse{\equal{}{#1}}%
{\mathbf{I}{\left[{#2}\vphantom{|_1^1}:\vphantom{|_1^1}{#3}\middle|{#4}\right]}}%
{\mathop{\mathbf{I}}_{#1}{\left[{#2}\vphantom{|_1^1}:\vphantom{|_1^1}{#3}\middle|{#4}\right]}}%
}

\providecommand{\E}[2][]{\mathop{\mathbf{E}}_{#1}{\lf[{#2}\rt]}}
\newcommand{\Ee}[3][]
{\mathop{\mathbf{E}}_{#1}{\lf[{#2}\vphantom{|_1^1}\md|\vphantom{|_1^1}{#3}\rt]}}

\newcommand{\Var}[1]{\mathop{\mathbf{Var}}{\left[{#1}\right]}}

\newcommand{\PR}[2][]{\mathop{\mathbf{Pr}}_{#1}{\left[{#2}\right]}}
\newcommand{\PRr}[3][]{\mathop{\mathbf{Pr}}_{#1}{\left[{#2}\vphantom{|_1^1}\md|\vphantom{|_1^1}{#3}\right]}}

\renewcommand{\U}[1][]{\ifthenelse{\equal{}{#1}}%
{{\cal U}}%
{{\cal U}_{#1}}}

\newcommand{\wt}{\widetilde}

\newcommand{\pss}[1][]{\nospell{\ifthenelse{\equal{}{#1}}%
{\txt{'s}}%
{\fla{#1\txt{'s}}}}}

\newcommand{\pl}[1][]{\nospell{\ifthenelse{\equal{}{#1}}%
{\mskip-6mu\stackrel{\text-}{}\mskip-4mu\txt{s}}%
{\fla{#1\mskip-6mu\stackrel{\text-}{}\mskip-4mu\txt{s}}}}}

\newcommand{\ord}[1][]{\nospell{\ifthenelse{\equal{}{#1}}%
{\txt{'th}}%
{\ifthenelse{\equal{1}{#1}}{$1\txt{'st}$}{\ifthenelse{\equal{2}{#1}}{$2\txt{'nd}$}{\ifthenelse{\equal{3}{#1}}{$3\txt{'rd}$}{\fla{#1\txt{'th}}}}}}}}

\newcommand{\fr}[3][*]{%
\ifthenelse{\equal{*}{#1}}%
{\frac{#2}{#3}}{}%
\ifthenelse{\equal{/}{#1}}%
{\nicefrac{#2}{#3}}{}%
\ifthenelse{\equal{}{#1}}%
{\left.#2\middle/#3\right.}{}%
\ifthenelse{\equal{p_}{#1}}%
{\left.\left(#2\right)\middle/#3\right.}{}%
\ifthenelse{\equal{_p}{#1}}%
{\left.#2\middle/\left(#3\right)\right.}{}%
\ifthenelse{\equal{pp}{#1}}%
{\left.\left(#2\right)\middle/\left(#3\right)\right.}{}%
}

\newcommand{\dr}{\nicefrac}

\newcommand{\sq}{\sqrt}

\newcommand{\set}[2][]{\ifthenelse{\equal{}{#1}}%
{\Ensuremath{\left\{#2\right\}}}%
{\Ensuremath{\left\{#2\vphantom{|_1^1}\md|\vphantom{|_1^1}#1\right\}}}}
\newcommand{\sett}[2]{\Ensuremath{\left\{#1\vphantom{|_1^1}\md|\vphantom{|_1^1}#2\right\}}}

\newcommand{\Min}[2][]{\ifthenelse{\equal{}{#1}}%
{\Ensuremath{\min{\left\{#2\right\}}}}%
{\Ensuremath{\min{\left\{#2\vphantom{|_1^1}\md|\vphantom{|_1^1}#1\right\}}}}}

\newcommand{\Inff}[3][]{\ifthenelse{\equal{}{#1}}%
{\Ensuremath{\inf_{#2}{\left\{#3\right\}}}}%
{\Ensuremath{\inf_{#2}{\left\{#3\vphantom{|_1^1}\md|\vphantom{|_1^1}#1\right\}}}}}

\newfunction{\asO}{O}
\newfunction{\asOm}{\Omega}
\newfunction{\asT}{\Theta}

\mydef{\01}{\set{0,1}}

\newcommand{\lf}{\left}
\newcommand{\rt}{\right}
\newcommand{\md}{\middle}

\newcommand{\sz}[2][]{\ifthenelse{\equal{}{#1}}%
{\Ensuremath{\left|#2\right|}}%
{\Ensuremath{\left|#2\right|_{#1}}}}

\newcommand{\sszz}[2]{\lf|\vphantom{|_1^1}\lf\{#1\md|#2\rt\}\vphantom{|_1^1}\rt|}

\newcommand{\txt}[1]{\textrm{#1}}  %

\newcommand{\tit}[1]{\textit{#1}}  %

\newcommand{\Cl}{\mathcal}  %

\DeclareMathAlphabet{\mathlowcal}{OT1}{pzc}{m}{it}

\newident[]{\R}{\mathcal R_{#1}}

\mydef{\l(}{\left(}
\mydef{\r)}{\right)}

\newmat{\mset}{\smin\set}

\newcommand{\nin}{\not\in}  %

\newcommand{\from}{\leftarrow}
\newcommand{\To}{\Rightarrow}
\newcommand{\Then}{\Longrightarrow}

\newcommand{\dt}{\cdot}
\newcommand{\tm}{\cdot}
\newcommand{\xor}{\oplus}
\newcommand{\sbseq}{\subseteq}
\newcommand{\sbs}{\subset}
\newcommand{\smin}{\setminus}
\mydef{\eps}{\varepsilon}
\newcommand{\deq}{\stackrel{\textrm{def}}{=}}
\newcommand{\unin}{\mathrel{\subset\mkern-13.1mu\sim}}  %

\providemat{\QQ}{\mathbb{Q}}
\providemat{\NN}{\mathbb{N}}
\providemat{\CC}{\mathbb{C}}
\providemat{\RR}{\mathbb{R}}
\providemat{\ZZ}{\mathbb{Z}}

\newcommand{\ds}[1][]
{\ifthenelse{\equal{}{#1}}{\allowbreak\dots}{#1\allowbreak\dots#1}}
\newmat{\dc}{\ds[,]}

\providecommand{\email}[1]{\url{mailto:#1}}

\newcommand{\abstr}[1]{\begin{abstract} #1 \end{abstract}}

{} %
\newcommand{\itemi}[2][*]{\ifthenelse{\equal{*}{#1}}%
{\begin{itemize}[noitemsep] #2 \end{itemize}}%
{\begin{itemize}[#1] #2 \end{itemize}}}

\newcommand{\itstart}[1][*]{\ifthenelse{\equal{*}{#1}}%
{\begin{itemize}[noitemsep]}%
{\begin{itemize}[#1]}%
}
\newcommand{\itend}{\end{itemize}}
{}  %

\protected \def \dg #1{%
\textcolor{Red}
{
{\normalmarginpar\marginnote{\bl{DG's comment}}}
{\reversemarginpar\marginnote{\bl{DG's comment}}\\}
\IfMathMode{
~~~\txt{#1}~
}{
~\\~~~#1~\\
{\normalmarginpar\marginnote{\bl{\ul{------}}}}
{\reversemarginpar\marginnote{\bl{\ul{------}}}\\}
}
}
\ClassWarning{My Macros}{#1}
}

\newcommand{\fn}[2][]{%
\IfMathMode{}{}%
\ifthenelse{\equal{}{#1}}%
{\footnote{#2}}%
{\footnote{\label{#1}#2}}%
}


{} %
\newcommand{\e}{\emph}
{}  %

\newcommand{\bl}[1]{{\bf #1}} %

\newcommand{\bil}[1]{{\bfseries\itshape #1}} %

\providecommand{\ul}[1]{\underline{#1}} %

\newcommand{\tb}{\quad}
\newcommand{\tbbb}{\qquad\qquad}

\makeatother%

{}

{}

\makeatother%

\title{On the randomised query complexity of composition}

\newcommand{\instDG}{Institute of Mathematics, Czech Academy of Sciences, \v Zitna 25, Praha 1, Czech Republic.}
\newcommand{\thanksDG}{Partially supported by the Grant No.\ P202/12/G061 of GA \v CR and by RVO:\ 67985840.
Part of this work was done while visiting the Centre for Quantum Technologies at the National University of Singapore.}

\newcommand{\instTL}{Division of Mathematical Sciences, Nanyang Technological University, Singapore and Centre for Quantum Technologies, National University of Singapore, Singapore. \email{troyjlee@gmail.com}.}
\newcommand{\ThanksTL}{}

\newcommand{\instMS}{IRIF, Universit\'e Paris Diderot, CNRS, 75205 Paris, France, and Centre for Quantum Technologies, National University of Singapore, Singapore. \email{santha@irif.fr}.}
\newcommand{\ThanksMS}{}

\newcommand{\thanksAll}{Partially supported by the National Research Foundation, including under NRF RF Award No.\ NRF-NRFF2013-13, the Prime Ministers Office, Singapore and the Ministry of Education, Singapore under the Research Centres of Excellence programme and by Grant No.\ MOE2012-T3-1-009.}

\author{
Dmitry Gavinsky\thanks{\instDG\ \thanksDG}~\thanks{\thanksAll}
\and Troy Lee\thanks{\instTL\ \ThanksTL}~\protect\footnotemark[2]
\and Miklos Santha\thanks{\instMS\ \ThanksMS}~\protect\footnotemark[2]
}

\begin{document}

\maketitle

\thispagestyle{empty}

\abstr{Let $f\sbseq\01^n\times\Xi$ be a relation and $g:\01^m\to\set{0,1,*}$ be a promise function.
This work investigates the randomised query complexity of the relation $f\circ g^n\sbseq\01^{m\tm n}\times\Xi$, which can be viewed as one of the most general cases of composition in the query model (letting $g$ be a relation seems to result in a rather unnatural definition of $f\circ g^n$).

We show that for every such $f$ and $g$,
\m{
\R(f\circ g^n) \in \asOm{\R(f)\tm\sq{\R(g)}}
,}
where $\R$ denotes the randomised query complexity.
On the other hand, we demonstrate a relation $f_0$ and a promise function $g_0$, such that $\R(f_0)\in\asT{\sq n}$, $\R(g_0)\in\asT{n}$ and $\R(f_0\circ g_0^n)\in\asT{n}$ -- that is, our composition statement is tight.

To the best of our knowledge, there was no known composition theorem for the randomised query complexity of relations or promise functions (and for the special case of total functions our lower bound gives multiplicative improvement of $\sq{\log n}$).
}

\sect[s_intro]{Introduction}

Let $f\sbseq\01^n\times\Xi$ be a relational problem over \f n-bit input strings, where $\Xi$ is a finite set and $\xi\in\Xi$ is a correct answer to $f(z)$ if and only if $(z,\xi)\in f$ -- in that case we write $\xi\in f(z)$, thus viewing $f(z)$ as the set of correct answers.
Let $g:\01^m\to\set{0,1,*}$ be a partial function over \f m-bit input strings, where ``$*$'' marks ``forbidden'' input strings -- in other words, $g$ is a Boolean promise function.
We will call $x\in\01^m$ a \e{legal} input value for $g$ if $g(x)\in\01$.

Define $f\circ g^n\sbseq\01^{n\tm m}\times\Xi$ as a relational problem over $n\tm m$-bit strings $x=(x_1\dc x_n)$, where
\m{
\twocase
{f\circ g^n(x)=\Xi}
{if $*\in\sett{g(x_i)}{i\in[n]}$;}
{f\circ g^n(x)=f(g(x_1)\dc g(x_n))}
{otherwise.}
}
That is, if at least one of \pl[x_i] as input to $g(\tm)$ violates the promise, then any $\xi\in\Xi$ is a correct answer to $f\circ g^n(x)$; otherwise, $f\circ g^n(x)$ is defined naturally.

We investigate the randomised query complexity ($\R$) of the relation $f\circ g^n$.
This setting can be viewed as one of the most general cases of so-called composition questions in the model of randomised query complexity:
Arguably, the most general natural way of modelling a computational problem in the query model is via a relation; on the other hand, letting the ``bottom'' problem $g$ be a relation seems to result in a rather awkward definition of the composed problem $f\circ g^n$ -- so, we've chosen to restrict $g$ by making it a promise function (the ``top'' problem $f$ may be a relation).\fn
{
Our lower bound argument would probably generalise to the case of both $f$ and $g$ being relations, though we haven't verified that.
}

We show that for \e{every} such $f$ and $g$,
\m{
\R(f\circ g^n) \in \asOm{\R(f)\tm\sq{\R(g)}}
.}

On the other hand, we demonstrate a relation $f_0$ and a promise function $g_0$, such that $\R(f_0)\in\asT{\sq n}$, $\R(g_0)\in\asT{n}$ and $\R(f_0\circ g_0^n)\in\asT{n}$ -- that is, our composition statement is tight.\fn
{
It may be worth noticing that the same example witnesses the possibility of $\R(f\circ g^n) \in \asO{\R(g)}$ when $\R(f)\in\asOm{\sq n}$.
}

\ssect*{Previous work}

To the best of our knowledge, prior to this work no general lower bound was known for the randomised query complexity of composed promise functions or relations.
For the special case of $g$ being a total function, Ben-David and Kothari~\cite{BK16_Ra_Q} have shown that 
\m{
\R(f\circ g^n) \in \asOm{\R(f)\tm\sq{\fr{\R(g)}{\log\l(\R(g)\r)}}}
.}

\ssect*{Our approach}

To argue that $\R(f\circ g^n) \in \asOm{\R(f)\tm\sq{\R(g)}}$, we will assume that a protocol for computing $f\circ g^n$ is given and use it to construct a protocol for computing $f$.
Our construction will be such that the new protocol will be accurate (with respect to $f$) if the given protocol was accurate (with respect to $f\circ g^n$), and the query complexity of the new protocol will be small if that of the given protocol was small.
As the query complexity of a protocol computing $f$ cannot be below $\R(f)$, a lower bound on the query complexity of the given protocol for $f\circ g^n$ will follow.

\sect[s_trees]{Polarised protocol trees}

In this part we describe a tree-like primitive for representing query protocols.
Some special properties of this representation will be useful for our analysis.

Let $f\sbseq\01^n\times\Xi$, $g:\01^m\to\set{0,1,*}$ and $f\circ g^n\sbseq\01^{n\tm m}\times\Xi$ be as described above.
Denote by $X=(X_1\dc X_n)$ the input to $f\circ g^n$, where every $X_i\in\01^m$ is input to $g$ (we will write $X_{i,j}$ to address the \ord[j] bit of $X_i$).
Denote by $Z\in\set{0,1,*}^n$ the input to $f$ -- in other words, $\forall i:~Z_i=g(X_i)$.

All protocol trees will have leaves, usually labelled with the answer returned by the protocol upon reaching this leaf.

A protocol for $f\circ g^n$ queries the values of $X_{i,j}$, so it will be represented by a tree, containing naturally defined nodes
\newcommand{\figXij}{$X_{i,j}$\,?}
\newcommand{\figO}{\fb 0}
\newcommand{\figI}{\fb 1}
\makeatletter%
\\\resizebox{.2\textwidth}{!}{%
\begin{picture}(0,0)%
\includegraphics{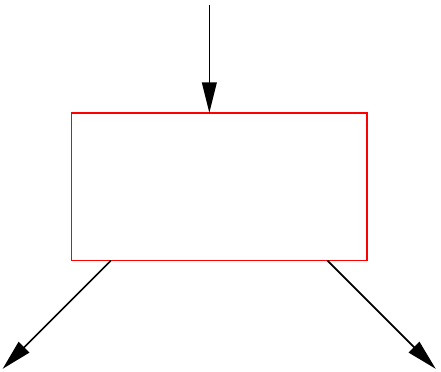}%
\end{picture}%
\setlength{\unitlength}{4144sp}%
\begingroup\makeatletter\ifx\SetFigFont\undefined%
\gdef\SetFigFont#1#2#3#4#5{%
\reset@font\fontsize{#1}{#2pt}%
\fontfamily{#3}\fontseries{#4}\fontshape{#5}%
\selectfont}%
\fi\endgroup%
\begin{picture}(2004,1689)(3454,-3358)
\put(4141,-2626){\makebox(0,0)[lb]{\smash{{\SetFigFont{20}{24.0}{\rmdefault}{\mddefault}{\updefault}{\color[rgb]{0,0,0}\figXij}%
}}}}
\put(3646,-3031){\makebox(0,0)[lb]{\smash{{\SetFigFont{12}{14.4}{\rmdefault}{\mddefault}{\updefault}{\color[rgb]{0,0,0}\figO}%
}}}}
\put(5176,-3031){\makebox(0,0)[lb]{\smash{{\SetFigFont{12}{14.4}{\rmdefault}{\mddefault}{\updefault}{\color[rgb]{0,0,0}\figI}%
}}}}
\end{picture}%%
}
\makeatother%

Leaves and \f X-queries are the only types of nodes that will occur in a tree representing a protocol for $f\circ g^n(X_1\dc X_n)$.
Note that the actions corresponding to these nodes are \e{deterministic}, and such will be our protocols for $f\circ g^n$.

Our protocols for $f$ will use \e{randomness},
represented by \e{randomised forks} in a tree:
\newcommand{\figa}{\fb{\alpha}}
\newcommand{\figIa}{\fb{1-\alpha}}
\makeatletter%
\\\resizebox{.2\textwidth}{!}{%
\begin{picture}(0,0)%
\includegraphics{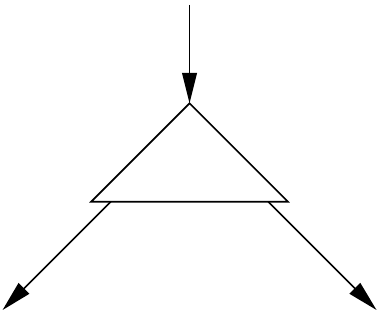}%
\end{picture}%
\setlength{\unitlength}{4144sp}%
\begingroup\makeatletter\ifx\SetFigFont\undefined%
\gdef\SetFigFont#1#2#3#4#5{%
\reset@font\fontsize{#1}{#2pt}%
\fontfamily{#3}\fontseries{#4}\fontshape{#5}%
\selectfont}%
\fi\endgroup%
\begin{picture}(1734,1419)(3679,-2278)
\put(5131,-1951){\makebox(0,0)[lb]{\smash{{\SetFigFont{12}{14.4}{\rmdefault}{\mddefault}{\updefault}{\color[rgb]{0,0,0}\figIa}%
}}}}
\put(3781,-1951){\makebox(0,0)[lb]{\smash{{\SetFigFont{12}{14.4}{\rmdefault}{\mddefault}{\updefault}{\color[rgb]{0,0,0}\figa}%
}}}}
\end{picture}%%
}\\
\makeatother%
Here the left son is selected with probability $\alpha$ and the right one with probability $1-\alpha$.

As we describe next, our protocol trees for $f$ will use \e{generic} nodes, somewhat non-standard:
Usually a vertex in a tree corresponds to certain state of the protocol and fully determines the ``history'' at that state -- namely, which queries have been made so far and what answers have been received from the oracle.
For reasons that will become clear later, our trees will have vertices corresponding to ``randomised \f Z-queries'', where certain $Z_i$ is queried with some probability between $0$ and $1$ -- accordingly, a pointer to a vertex coming after such ``uncertain query'' does not reveal whether $Z_i$ has actually been queried.
To address this, we accompany our trees for computing $f(Z)$ with ``memory'' $w\in\set{0,1,*}^n$, such that $w_i=Z_i$ if $Z_i$ has already been queried by the protocol (in which case we will always assume $Z_i\in\01$, as explained next) and $w_i=*$ otherwise.

Our trees will have two types of generic vertices, corresponding to certain complex protocol's actions.
The reason why we prefer to treat these complex actions as single tree nodes is the following:
Recall that we will construct a protocol for $f$, based on a (given) protocol for $f\circ g^n$; each generic vertex in the new protocol will correspond to a single $X_{i,j}$-query in the original protocol; as a result, the tree of the constructed protocol for $f$ will be \e{isomorphic} to the tree of the given protocol for $f\circ g^n$.
Allowing generic vertices and using the auxiliary registers $w_i$ are the ``price'' of keeping this very convenient isomorphism between the original and the constructed protocol trees.

The first generic type is a \e{\f Z-node}, described by $(i,\alpha,\beta)\in[n]\times[0,1]\times[0,1]$:
\newcommand{\figZi}{$Z_i$\,?}
\newcommand{\figwi}{$w_i$\,?}
\newcommand{\figwiZi}{\ul{$w_i\from Z_i$}}
\newcommand{\figwist}{Is $w_i=*$?}
\newcommand{\figyes}{\bl{yes}}
\newcommand{\figno}{\bl{no}}
\newcommand{\figb}{\fb{\beta}}
\newcommand{\figIb}{\fb{1-\beta}}
\newcommand{\figOp}{``\fb 0''}
\newcommand{\figIp}{``\fb 1''}
\makeatletter%
\\\resizebox{.4\textwidth}{!}{%
\begin{picture}(0,0)%
\includegraphics{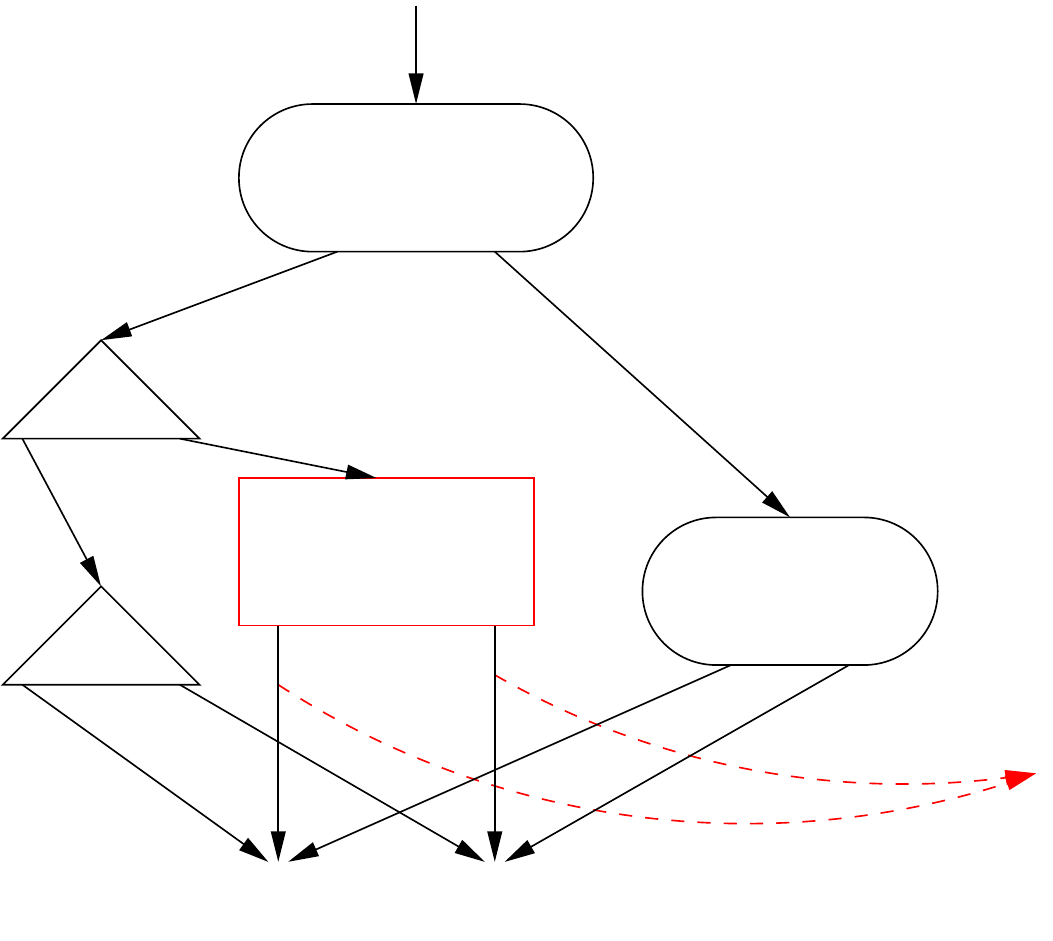}%
\end{picture}%
\setlength{\unitlength}{4144sp}%
\begingroup\makeatletter\ifx\SetFigFont\undefined%
\gdef\SetFigFont#1#2#3#4#5{%
\reset@font\fontsize{#1}{#2pt}%
\fontfamily{#3}\fontseries{#4}\fontshape{#5}%
\selectfont}%
\fi\endgroup%
\begin{picture}(4797,4288)(2689,-4112)
\put(4186,-2446){\makebox(0,0)[lb]{\smash{{\SetFigFont{20}{24.0}{\rmdefault}{\mddefault}{\updefault}{\color[rgb]{0,0,0}\figZi}%
}}}}
\put(6031,-2626){\makebox(0,0)[lb]{\smash{{\SetFigFont{20}{24.0}{\rmdefault}{\mddefault}{\updefault}{\color[rgb]{0,0,0}\figwi}%
}}}}
\put(3961,-736){\makebox(0,0)[lb]{\smash{{\SetFigFont{20}{24.0}{\rmdefault}{\mddefault}{\updefault}{\color[rgb]{0,0,0}\figwist}%
}}}}
\put(5581,-2941){\makebox(0,0)[lb]{\smash{{\SetFigFont{12}{14.4}{\rmdefault}{\mddefault}{\updefault}{\color[rgb]{0,0,0}\figO}%
}}}}
\put(6481,-3076){\makebox(0,0)[lb]{\smash{{\SetFigFont{12}{14.4}{\rmdefault}{\mddefault}{\updefault}{\color[rgb]{0,0,0}\figI}%
}}}}
\put(4006,-2851){\makebox(0,0)[lb]{\smash{{\SetFigFont{12}{14.4}{\rmdefault}{\mddefault}{\updefault}{\color[rgb]{0,0,0}\figO}%
}}}}
\put(4996,-2851){\makebox(0,0)[lb]{\smash{{\SetFigFont{12}{14.4}{\rmdefault}{\mddefault}{\updefault}{\color[rgb]{0,0,0}\figI}%
}}}}
\put(7471,-3436){\makebox(0,0)[lb]{\smash{{\SetFigFont{20}{24.0}{\rmdefault}{\mddefault}{\updefault}{\color[rgb]{1,0,0}\figwiZi}%
}}}}
\put(3736,-1276){\makebox(0,0)[lb]{\smash{{\SetFigFont{12}{14.4}{\rmdefault}{\mddefault}{\updefault}{\color[rgb]{0,0,0}\figyes}%
}}}}
\put(5356,-1276){\makebox(0,0)[lb]{\smash{{\SetFigFont{12}{14.4}{\rmdefault}{\mddefault}{\updefault}{\color[rgb]{0,0,0}\figno}%
}}}}
\put(2926,-2041){\makebox(0,0)[lb]{\smash{{\SetFigFont{12}{14.4}{\rmdefault}{\mddefault}{\updefault}{\color[rgb]{0,0,0}\figIa}%
}}}}
\put(3826,-1861){\makebox(0,0)[lb]{\smash{{\SetFigFont{12}{14.4}{\rmdefault}{\mddefault}{\updefault}{\color[rgb]{0,0,0}\figa}%
}}}}
\put(3466,-3166){\makebox(0,0)[lb]{\smash{{\SetFigFont{12}{14.4}{\rmdefault}{\mddefault}{\updefault}{\color[rgb]{0,0,0}\figb}%
}}}}
\put(2836,-3481){\makebox(0,0)[lb]{\smash{{\SetFigFont{12}{14.4}{\rmdefault}{\mddefault}{\updefault}{\color[rgb]{0,0,0}\figIb}%
}}}}
\put(4771,-4021){\makebox(0,0)[lb]{\smash{{\SetFigFont{14}{16.8}{\rmdefault}{\mddefault}{\updefault}{\color[rgb]{0,0,0}\figIp}%
}}}}
\put(3781,-4021){\makebox(0,0)[lb]{\smash{{\SetFigFont{14}{16.8}{\rmdefault}{\mddefault}{\updefault}{\color[rgb]{0,0,0}\figOp}%
}}}}
\end{picture}%%
}\\
\makeatother%
In words, this generic node corresponds to the following action:\itemi{
\item If $w_i=*$, then with probability $\alpha$ a query is made to read the value of $Z_i$ and one of the two outgoing edges is selected accordingly; otherwise (i.e., with probability $1-\alpha$), the outgoing edge marked with ``$1$'' is selected with probability $\beta$ and the edge marked with ``$0$'' is selected otherwise (i.e., with probability $1-\beta$).
\item If $w_i\neq*$, then the outgoing edge is selected according to the (already known) value $Z_i=w_i$.
\item If a query to $Z_i$ has been made (which only occurs with probability $\alpha$ when $w_i=*$), then the value of the register $w_i$ is updated to contain the value of $Z_i$.
}

Observe that in the above description we have assumed that if a query to $Z_i$ has been made (i.e., $w_i\neq*$), then $Z_i\in\01$ -- namely, $X_i$ is a legal input value for $g$.
We will be making this convenient assumption (sometimes implicitly) throughout our lower-bound analysis, as our target distributions are always supported on legal input values only.

The second generic type is a \e{\f Z-mixer}, described by $(i,\alpha,\beta)\in[n]\times[0,1]\times[0,1]$ and acting as follows:
\makeatletter%
\\\resizebox{.3\textwidth}{!}{%
\begin{picture}(0,0)%
\includegraphics{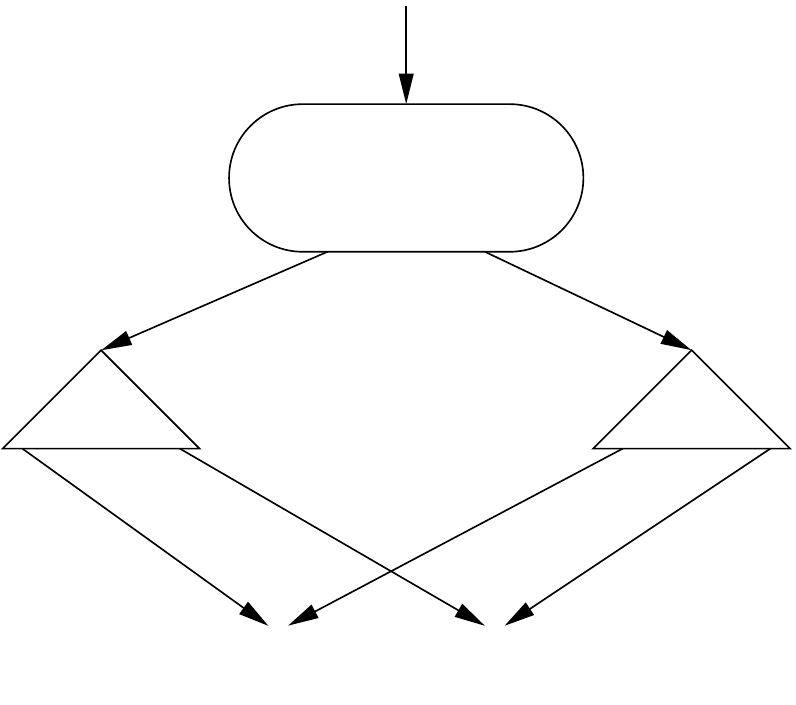}%
\end{picture}%
\setlength{\unitlength}{4144sp}%
\begingroup\makeatletter\ifx\SetFigFont\undefined%
\gdef\SetFigFont#1#2#3#4#5{%
\reset@font\fontsize{#1}{#2pt}%
\fontfamily{#3}\fontseries{#4}\fontshape{#5}%
\selectfont}%
\fi\endgroup%
\begin{picture}(3624,3208)(2689,-4112)
\put(3466,-3166){\makebox(0,0)[lb]{\smash{{\SetFigFont{12}{14.4}{\rmdefault}{\mddefault}{\updefault}{\color[rgb]{0,0,0}\figb}%
}}}}
\put(2836,-3481){\makebox(0,0)[lb]{\smash{{\SetFigFont{12}{14.4}{\rmdefault}{\mddefault}{\updefault}{\color[rgb]{0,0,0}\figIb}%
}}}}
\put(4771,-4021){\makebox(0,0)[lb]{\smash{{\SetFigFont{14}{16.8}{\rmdefault}{\mddefault}{\updefault}{\color[rgb]{0,0,0}\figIp}%
}}}}
\put(3781,-4021){\makebox(0,0)[lb]{\smash{{\SetFigFont{14}{16.8}{\rmdefault}{\mddefault}{\updefault}{\color[rgb]{0,0,0}\figOp}%
}}}}
\put(5896,-3301){\makebox(0,0)[lb]{\smash{{\SetFigFont{12}{14.4}{\rmdefault}{\mddefault}{\updefault}{\color[rgb]{0,0,0}\figa}%
}}}}
\put(4771,-3076){\makebox(0,0)[lb]{\smash{{\SetFigFont{12}{14.4}{\rmdefault}{\mddefault}{\updefault}{\color[rgb]{0,0,0}\figIa}%
}}}}
\put(3916,-1816){\makebox(0,0)[lb]{\smash{{\SetFigFont{20}{24.0}{\rmdefault}{\mddefault}{\updefault}{\color[rgb]{0,0,0}\figwist}%
}}}}
\put(4996,-2356){\makebox(0,0)[lb]{\smash{{\SetFigFont{12}{14.4}{\rmdefault}{\mddefault}{\updefault}{\color[rgb]{0,0,0}\figno}%
}}}}
\put(3736,-2356){\makebox(0,0)[lb]{\smash{{\SetFigFont{12}{14.4}{\rmdefault}{\mddefault}{\updefault}{\color[rgb]{0,0,0}\figyes}%
}}}}
\end{picture}%%
}\\
\makeatother%
In words, this generic node corresponds to the following action:\itemi{
\item If $w_i=*$, then the outgoing edge marked with ``$1$'' is selected with probability $\beta$ and the edge marked with ``$0$'' is selected otherwise -- i.e., with probability $1-\beta$.
\item If $w_i\neq*$, then the outgoing edge marked with ``$1$'' is selected with probability $\alpha$ and the edge marked with ``$0$'' is selected otherwise -- i.e., with probability $1-\alpha$.
}

\ssect{Some properties of our trees}

When constructing a protocol for computing $f$, we will try to keep low the (expected) number of actual \f Z-queries made by the protocol.
Note that the only node type where $Z_i$ may be queried are the \f Z-nodes.
There a query to $Z_i$ can only take place if $w_i=*$, and when that happens the value of $w_i$ is updated -- therefore, \e{each $Z_i$ can be queried at most once}.

The other property of interest to us is also related to ``saving'' \f Z-queries.
We call it \e{polarity}, it says that (for all $i\in[n]$) the set of computational paths\fn
{
When a tree contains generic nodes, by a \e{computational path} we mean not only the set of tree nodes that have been ``visited'', but also the information about the randomised decisions taken at each visited generic node.
Accordingly, a number of distinct computational paths can lead to the same tree leaf.
}
leading to the same tree vertex can either consist of \e{paths that have not queried $Z_i$ and those where $Z_i=0$} or of \e{paths that have not queried $Z_i$ and of those where $Z_i=1$} -- in other words, \e{a path ``knowing'' that $Z_i=0$ and a path ``knowing'' that $Z_i=1$ cannot lead to the same vertex} in our tree.
That is why we are calling our trees \e{polarised}.

To see that they are indeed polarised, note that only the two generic node types have merging paths; \f Z-mixers never make queries, and therefore cannot affect the polarity; a \f Z-node can only merge paths of the ``same polarity'' (if $w_i=1$, then the corresponding paths can only mix with the \f1-outcome of the \f Z-query, and vice versa), thus preserving the polarity.

Informally, the reason why polarity will be important for us is this.
Suppose that a tree were not polarised and let $l_0$ be a leaf, such that exactly half of the paths leading to it ``know'' that $Z_i=0$, while the other half ``know'' that $Z_i=1$.
On the one hand, the ``leaf-wise'' knowledge in $l_0$ about the value of $Z_i$ is the lowest possible (the entropy of the bit $Z_i$, conditioned on reaching $l_0$ is $1$); on the other, the probability that $Z_i$ has actually been queried by the protocol when $l_0$ is reached is $1$ -- so, conditional on reaching $l_0$, a query to $Z_i$ has been fully ``wasted'' (as a leaf, $l_0$ must correspond to an answer to $f(Z_1\dc Z_n)$, and this answer must be fully independent of $Z_i$).

By having chosen carefully the set of allowed generic actions in a tree, we are guaranteeing that it is ``ontologically polarised'' -- thus avoiding the possibility of ``wasted queries'', as described above.
On the other hand, having such ``reduced instruction set'' will demand from us a somewhat bigger effort in order to ``mimic'' the behaviour of the given protocol for $f\circ g^n$ when constructing a protocol for $f$.
We will see next how to achieve that; as a result, the constructed protocol will compute $f(Z)$ with exactly the same accuracy that the original protocol achieves for $f\circ g^n(X_1\dc X_n)$.

\sect[s_prot]{Constructing a protocol for $f$, given a protocol for $f\circ g^n$}

Recall that the argument of our lower bound is based on transforming a given protocol for $f\circ g^n$ into a protocol for $f$, as accurate as the original one and whose query complexity will be low if that of the original protocol was low.
In this part we describe the transformation and notice some basic properties of the constructed protocol for $f$, as summarised by \lemref{l_P'} and \crlref{crl_P'} (used in \sref{s_compl} to obtain the desired bound).

Let $\mu_g$ over $\01^m$ be a non-trivial\fn
{
so that $g$ is not constant on $\supp(\mu_g)$
}
input distribution for $g$ that is supported on legal input values only, and for $a\in\01$ denote by $\mu_g^a$ the distribution of $Y\in\01^m$ when $Y\sim\mu_g$, conditioned on $g(Y)=a$.
For a binary string $z$ of length $k$, denote by $z\circ\mu_g$ the distribution of $(X_1\dc X_k)$, where $X_i\sim\mu_g^{z_i}$ for every $i\in[k]$.
For a distribution $\nu$ on $\01^k$, denote by
\m{\nu\circ\mu_g}
the distribution of $X\in\01^{k\tm m}$, corresponding to choosing $Z\sim\nu$ followed by $X\sim Z\circ\mu_g$.

Let $\mu$ over $\01^{n\tm m}$ be input distribution for $f\circ g^n$, so that $\mu=\mu_f\circ\mu_g$ for some $\mu_f$ over $\01^n$ -- in other words, $\mu_f$ is the distribution of $(g(X_1)\dc g(X_n))$ when $X=(X_{1}\dc X_{n})\sim\mu$.

Let $\Cl P$ be a (deterministic) protocol that solves $f\circ g^n$ with respect to $\mu$ with error $\eps$.
Next we construct a protocol $\Cl P'$ that solves $f$ with respect to $\mu_f$ with the same error $\eps$.
We will represent $\Cl P'$ as a polarised protocol tree, which will be isomorphic to the tree of $\Cl P$.

\ssect[ss_step]{Constructing $\Cl P'$: the inductive step}

Let $Z$ be a random variable taking values in $\01^n$:\ in the context of computing $f\circ g^n(X_1\dc X_n)$ we let $Z_i=g(X_i)$, and in the context of computing $f$ we let $Z_i$ be the \ord[i] bit of input (this convention should not confuse us, as $Z$ will denote the input to $f$ in both cases).

To build a protocol for $f$, we repeatedly apply the following ``local'' mechanism that ``translates'' every node of the (given) tree for $f\circ g^n$ into a node of a polarised tree for $f$.

Starting from the root and selecting at each step a new non-leaf vertex whose predecessor in $\Cl P'$ has already been constructed, we proceed as follows.
Denote by $v_0$ be the current node in $\Cl P$ and by $v_0'$ the corresponding node in $\Cl P'$.
The action of $v_0$ is an \f X-query -- let it query $X_{i_0,j_0}$.
Let
\m{
p_{in}\deq\PRr[X\sim\mu]{Z_{i_0}=1}{v_0}
,}
where $[v_0]$ denotes the event that execution of $\Cl P$ has reached the node $v_0$.
Also let
\m{
p_<\deq\PRr[\mu]{Z_{i_0}=1}{[v_0], X_{i_0,j_0}=a_0}
\txt{\tb and\tb}
p_>\deq\PRr[\mu]{Z_{i_0}=1}{[v_0], X_{i_0,j_0}=1-a_0}
,}
where $a_0\in\01$ is such that $p_<\le p_>$, and
\m{
\tau_<\deq\PRr[\mu]{X_{i_0,j_0}=a_0}{v_0}
\txt{\tb and\tb}
\tau_>\deq\PRr[\mu]{X_{i_0,j_0}=1-a_0}{v_0}\tb\lf\{=1-\tau_<\rt\}
.}

The action of $v_0'$ will be either a \f Z-node or a \f Z-mixer.
We will associate the edge that leaves $v_0'$ and is marked by ``$0$'' with the ``$a_0$'' edge of $v_0$, and the edge that leaves $v_0'$ and is marked by ``$1$'' with the ``$1-a_0$'' edge of $v_0$.

Informally, we want the action of $v_0'$ to ``mimic'' that of $v_0$ with respect to the (conditional) distribution of $Z$: if we do that for every node in $\Cl P$, then in the end we will obtain a protocol that solves $f(Z)$ with respect to $\mu_f$ with the same accuracy as $\Cl P$ achieves for $f\circ g^n(X_1\dc X_n)$ with respect to $\mu$.
In order to imitate the behaviour of $\Cl P$ with respect to $Z$, the new protocol must (at least) imitate it at every $v_0$ with respect to the corresponding $Z_{i_0}$.

Technically, our assignment of action to $v_0'$ will be such that if
\m[m_indas]{
\PRr[\mu_f]{Z_{i_0}=1}{v_0'} = p_{in}
,}
then
\m[m_req]{
\PRr[\mu_f]{Z_{i_0}=1}{v_0',\txt{answer ``$0$''}} = p_<
\txt{\tb and\tb}
\PRr[\mu_f]{Z_{i_0}=1}{v_0',\txt{answer ``$1$''}} = p_>
}
and
\m[m_reqt]{
\PRr[\mu_f]{\txt{answer ``$0$''}}{v_0'} = \tau_<
\txt{\tb and\tb}
\PRr[\mu_f]{\txt{answer ``$1$''}}{v_0'} = \tau_>
,}
where $[\txt{answer ``$b$''}]$ is the event that the generic action assigned to $v_0'$ returns the corresponding answer.

Note that \bref{m_indas} and \bref{m_req} implies
\mal{
\tau_<\tm p_< + (1-\tau_<)\tm p_>
&= p_{in}\\
&= \PRr[\mu_f]{\txt{answer ``$0$''}}{v_0'}\tm p_<
+ \l(1-\PRr[\mu_f]{\txt{answer ``$0$''}}{v_0'}\r)\tm p_>
,}
and therefore \bref{m_reqt} ``almost always'':\ the only exception is the ``degenerate'' case $[p_<=p_>]$ (which we will handle soon).
In all other cases it will be enough to guarantee \bref{m_req}, assuming \bref{m_indas}, and that will imply \bref{m_reqt} as well.

Recall that all the predecessors of $v_0'$ in $\Cl P'$ have already been constructed, so the value
\m{
q_{in}\deq\PRr[\mu_f]{w_{i_0}\neq*}{v_0'}
}
is well-defined.
Observe that
\m{
p_<\le p_{in}\le p_>\le1
,}
as $\tau_<\tm p_<+\tau_>\tm p_>=p_{in}$ and $\tau_<+\tau_>=1$.

For the same reason, if $p_<=p_{in}$ or $p_{in}=p_>$, then $p_<=p_{in}=p_>$, which is the degenerate case mentioned above.
To handle it, we let the action of $v_0'$ be a \f Z-mixer $(i_0,\tau_>,\tau_>)$ -- it is easy to see that this choice satisfies both \bref{m_req} and \bref{m_reqt}.

Now assume that
\m[m_pina]{
p_<<p_{in}<p_>
.}
As our tree is polarised, either $\PRr{w_{i_0}=0}{v_0'}=q_{in}$ or $\PRr{w_{i_0}=1}{v_0'}=q_{in}$ holds -- without loss of generality, let us assume the latter (the other case is symmetric and treated similarly).

Let
\m{
p_{in}^*\deq\PRr{Z_{i_0}=1}{[v_0'],w_{i_0}=*}
}
and observe that
\m[m_eq]{
p_{in}^*\tm(1-q_{in}) + q_{in}
\,=\,p_{in}
\,=\,p_<\tm\tau_< + p_>\tm\tau_>
\,=\, p_<\tm\tau_< + p_>\tm(1-\tau_<)
}
-- these equalities will play their crucial role soon.

Note that the first equality in \bref{m_eq} implies
\m[m_pin]{
p_{in}^*=\fr{p_{in}-q_{in}}{1-q_{in}} \le p_{in}
,}
and so -- by assumption \bref{m_pina} -- either $p_{in}^*\le p_<<p_>$ or $p_<<p_{in}^*<p_>$ holds.
Both possibilities are valid and we will handle them differently:\ either with a \f Z-mixer or with a \f Z-node.

\sssect{The case of $p_{in}^*\le p_<$: using a \f Z-mixer}

Here we are assuming that
\m[m_psina]{
p_{in}^*\le p_<<p_>\le1
.}
We will choose such $\alpha_0,\beta_0\in[0,1]$ that making a \f Z-mixer $(i_0,\alpha_0,\beta_0)$ to be the action of $v_0'$ will satisfy \bref{m_req}.

Let $\gamma_1$ be such that
\m[m_p1]{
p_1(\gamma_1)\deq
\fr{p_{in}^*\tm(1-q_{in}) + \gamma_1}{1-q_{in}+\gamma_1}=p_<
,}
which exists and satisfies
\m[m_g1]{
\gamma_1\in[0,q_{in}]
,}
as $p_1(0)=p_{in}^*\le p_<$ by \bref{m_psina}, $p_1(q_{in})=p_{in}>p_<$ by \bref{m_eq} and \bref{m_pina}, and $p_1(\dt)$ is monotone on $[0,q_{in}]$, obviously.
Then
\mal{
p_<\tm(1-q_{in}+\gamma_1) + q_{in} - \gamma_1
&= p_{in}^*\tm(1-q_{in}) + q_{in}
= p_{in}\\
&=p_<\tm\tau_< + p_>\tm(1-\tau_<)
,}
where the first equality is \bref{m_p1} and the last two are \bref{m_eq}.
Since $p_>\le1$ and $\gamma_1\le q_{in}$,
\m{
p_<\tm(1-q_{in}+\gamma_1) + p_>\tm(q_{in}-\gamma_1)
\le p_<\tm\tau_< + p_>\tm(1-\tau_<)
}
and
\m{
p_<\tm\l(1-q_{in}+\gamma_1-\tau_<\r)
\le p_>\tm\l(1-q_{in}+\gamma_1-\tau_<\r)
}
-- that is,
\m[m_qin]{
1-q_{in}+\gamma_1\ge\tau_<
.}

Let $\alpha_0,\beta_0\in[0,1]$ be such that
\m{
1-\beta_0 = \fr{\tau_<}{1-q_{in}+\gamma_1}
}
and
\m{
(1-\alpha_0)\tm q_{in} = (1-\beta_0)\tm\gamma_1,
}
their existence follows from \bref{m_qin} and \bref{m_g1}, respectively.
We set the action of $v_0'$ to be a \f Z-mixer parametrised by $(i_0,\alpha_0,\beta_0)$ and we claim that \bref{m_req} is satisfied.

On the one hand,
\mal[P]{
\PRr[\mu_f]{Z_{i_0}=1}{v_0',\txt{answer ``$0$''}}
&=\fr{(1-q_{in})\tm(1-\beta_0)\tm p_{in}^* + q_{in}\tm(1-\alpha_0)\tm1}
{(1-q_{in})\tm(1-\beta_0) + q_{in}\tm(1-\alpha_0)}\\
&=\fr{(1-q_{in})\tm(1-\beta_0)\tm p_{in}^* + (1-\beta_0)\tm\gamma_1}
{(1-q_{in})\tm(1-\beta_0) + (1-\beta_0)\tm\gamma_1}\\
&=\fr{(1-q_{in})\tm p_{in}^* + \gamma_1}
{(1-q_{in}) + \gamma_1}
~=~p_<
,}
where the last equality is \bref{m_p1}.
On the other,
\mal{
\PRr[\mu_f]{\txt{answer ``$0$''}}{v_0'}
&=(1-q_{in})\tm(1-\beta_0)+q_{in}\tm(1-\alpha_0)\\
&=(1-q_{in}+\gamma_1)\tm(1-\beta_0)
~=~\tau_<
.}
Therefore,
\m{
\PRr[\mu_f]{\txt{answer ``$1$''}}{v_0'}
=1-\PRr[\mu_f]{\txt{answer ``$0$''}}{v_0'}
=1-\tau_<=\tau_>
}
and
\mal{
\PRr[\mu_f]{Z_{i_0}=1}{v_0',\txt{answer ``$1$''}}
&=\fr{p_{in}-\PRr{Z_{i_0}=1}{v_0',\txt{answer ``$0$''}}
\tm\PRr{\txt{answer ``$0$''}}{v_0'}}
{\PRr{\txt{answer ``$1$''}}{v_0'}}\\
&=\fr{p_{in}-p_<\tm\tau_<}{\tau_>}
=p_>
,}
as required.

\sssect{The case of $p_<<p_{in}^*$: using a \f Z-node}

Now assume that
\m{
p_<<p_{in}^*<p_>\le1
.}
We will choose such $\alpha_0,\beta_0\in[0,1]$ that making a \f Z-node $(i_0,\alpha_0,\beta_0)$ to be the action of $v_0'$ will satisfy \bref{m_req}.

Let $\alpha'\in[0,1]$ be such that
\m[m_alpha']{
\fr{(1-\alpha')\tm p_{in}^*}{1-p_{in}^*}
=\fr{p_<}{1-p_<}
,}
and denote
\m{
\gamma_2 \deq (1-\alpha'p_{in}^*)(1-q_{in})
\txt{\tb and\tb}
\gamma_3 \deq
\fr{\tau_<}{\gamma_2}
.}
From \bref{m_alpha'}:
\m[m_p<]{
p_<=\fr{(1-\alpha')\tm p_{in}^*}{1-\alpha'p_{in}^*}
,}
and so,
\mal[P]{
p_<\tm\gamma_2 + 1-\gamma_2
&=\fr{(1-\alpha')\tm p_{in}^*}{1-\alpha'p_{in}^*}\tm\gamma_2 + 1-\gamma_2\\
&=(1-\alpha')\tm p_{in}^*\tm(1-q_{in})
+1-(1-\alpha'p_{in}^*)(1-q_{in})\\
&=p_{in}^*\tm(1-q_{in}) + q_{in}\\
&=~p_{in}\\
&=~p_<\tm\tau_< + p_>\tm(1-\tau_<)
,}
where the last two equalities follow, respectively, from \bref{m_pin} and \bref{m_eq}.
As $p_>,\tau_<\in[0,1]$,
\m{
p_<\tm\gamma_2 + 1-\gamma_2 \le p_<\tm\tau_< + 1-\tau_<
,}
and it holds that $\gamma_2\ge\tau_<$ and
\m[m_g3]{
\gamma_3 \in[0,1]
.}

Let
\m{
\alpha_0 = \gamma_3 \tm \alpha'
\txt{\tb and\tb}
\beta_0 = \fr{1-\gamma_3}{1-\gamma_3 \tm \alpha'}
.}
It follows from $\alpha'\in[0,1]$ and \bref{m_g3} that $\alpha_0,\beta_0\in[0,1]$.
We set the action of $v_0'$ to be a \f Z-node parametrised by $(i_0,\alpha_0,\beta_0)$ and claim that \bref{m_req} is satisfied.

On the one hand,
\mal{
\PRr[\mu_f]{\txt{answer ``$0$''}}{v_0'}
&=(1-q_{in})
\tm\big( (1-\alpha_0)\tm(1-\beta_0) + \alpha_0\tm(1-p_{in}^*) \big)\\
&=(1-q_{in})\tm\big( 1-\alpha_0\tm p_{in}^*-\beta_0\tm(1-\alpha_0) \big)\\
&=(1-q_{in})\tm(\gamma_3-\gamma_3\tm \alpha'\tm p_{in}^*)\\
&=\tau_<\tm\fr{(1-q_{in})\tm(1-\alpha'\tm p_{in}^*)}{\gamma_2}
~=~\tau_<
;}
on the other,
\mal{
\PRr[\mu_f]{Z_{i_0}=1}{v_0',\txt{answer ``$0$''}}
&=\fr{(1-\alpha_0)\tm(1-\beta_0)\tm p_{in}^*}
{(1-\alpha_0)\tm(1-\beta_0) + \alpha_0\tm(1-p_{in}^*)}\\
&=\fr{(\gamma_3-\alpha_0)\tm p_{in}^*}
{\gamma_3-\alpha_0\tm p_{in}^*}\\
&=\fr{(1-\alpha')\tm p_{in}^*}
{1-\alpha'\tm p_{in}^*}
~=~p_<
,}
where the last equality is \bref{m_p<}.
As in the case of $p_{in}^*\le p_<$, from here it follows that \bref{m_req} holds.

\ssect[ss_prot]{Constructing $\Cl P'$: summing up}

Thus far, we have assigned actions to the internal nodes in a polarised tree representing $\Cl P'$, such that for every pair of mutually-corresponding non-leaves $v_0\in\Cl P$ and $v_0'\in\Cl P'$ that satisfy \bref{m_indas}, both \bref{m_req} and \bref{m_reqt} must hold.
Intuitively, it says that if the distribution of $Z_{i_0}$ was ``correct'' upon reaching $v_0'$, then it is stays ``correct'' also after the operation performed by $v_0'$, where ``correct'' means being the same as in $\Cl P$.

We would like to use this fact inductively in order to conclude that for every mutually-corresponding pair of vertices $v_0'\in\Cl P'$ and $v_0\in\Cl P$, the distribution of $Z\sim\mu_f$ conditioned on reaching $v_0'$ is the same as the distribution of $Z=g^n(X)$ for $X\sim\mu$ conditioned on reaching $v_0$.
For that we need a somewhat stronger ``step statement'' than what we have above:\ namely, we would like to say that if the distribution of the whole random vector $Z$ is ``correct'' upon reaching $v_0'$, then it stays ``correct'' after leaving $v_0'$ as well.

In other words, we want to argue that it is enough to imitate by the action of $v_0'$ the behaviour of $v_0$ with respect to the ``queried'' coordinate $Z_{i_0}$ (as we have done) in order to conclude that the behaviour with respect to full $Z$ has been mimicked as well.

First we claim that the distribution of $X$, conditioned on reaching a certain node in $\Cl P$, is always ``reasonably localised'' when $X\sim\mu$ (which holds due to the fact that $\mu$ has the ``concatenated'' structure of $\mu_f\circ\mu_g$).

\clm[cl_invP]{Let $v$ be a vertex in $\Cl P$ and $i\in[n]$, then
\m{
\Ii[X\sim\mu]{X_i}{X_{[n]\mset i}}{[v], Z_i} = 0.~\footnotemark
}
\footnotetext
{
Here ``$[v]$'' denotes the \e{event} that $v$ is reached by the protocol and ``$Z_i$'' stands for conditioning on the \e{value} that the variable takes.
}
}

Note that in general $\Ii{X_i}{X_{[n]\mset i}}{[v]}$ can be positive, as $\mu_f$ doesn't need to be a product distribution; however, in the above statement this possible dependence is ``shielded'' by conditioning on $Z_i$.

\prfstart
Let $a\in\01$ and recall that $\mu=\mu_f\circ\mu_g$.
By the nature of a query protocol, there exist sets $A$ and $B$ such that the distribution of $X\sim\mu$, conditioned on reaching $v$, is the same as the distribution of $X\sim\mu$, conditioned on $X_i\in A$ and $X_{[n]\mset i}\in B$.
Accordingly,
\m{
\Ii[X\sim\mu]{X_i}{X_{[n]\mset i}}{[v], Z_i=a}
&= \Ii[\mu]{X_i}{X_{[n]\mset i}}{Z_i=a, X_i\in A, X_{[n]\mset i}\in B}\\
&= \Ii[\mu']{X_i}{X_{[n]\mset i}}{X_i\in A, X_{[n]\mset i}\in B}
,}
where $\mu'=\mu_f'\circ\mu_g$ for $Z\sim\mu_f'$ defined as $Z\sim\mu_f$, subject to $[Z_i=a]$.
By our definition of concatenated distributions, $\I[\mu']{X_i}{X_{[n]\mset i}}=0$, and therefore
\m{
\Ii[\mu']{X_i}{X_{[n]\mset i}}{X_i\in A, X_{[n]\mset i}\in B} = 0
}
as well.

Finally, $\Ii[\mu]{X_i}{X_{[n]\mset i}}{[v], Z_i}$ is a convex combination of $\Ii[\mu]{X_i}{X_{[n]\mset i}}{[v], Z_i=0}$ and $\Ii[\mu]{X_i}{X_{[n]\mset i}}{[v], Z_i=1}$, as $\mu_g$ is supported on legal input values for $g$.
The result follows.
\prfend[\clmref{cl_invP}]

Let $v_0$ be a node in $\Cl P$ that queries $X_{i_0,j_0}$ and let $v_0'$ be the corresponding node in $\Cl P'$.
Assume
\m[m_alindas]{
\forall z\in\01^n:\,\PRr[Z\sim\mu_f]{Z=z}{v_0'} = \PRr[X\sim\mu]{Z=z}{v_0}
}
-- in particular, this means that \bref{m_indas} is satisfied, and therefore \bref{m_req} and \bref{m_reqt} hold with respect to $v_0$ and $v_0'$.

Let us see what happens at the vertices $v_0\in\Cl P$ and $v_0'\in\Cl P'$, conditioned upon $[Z_{i_0}=1]$.
First of all, from \bref{m_req} and \bref{m_reqt} it follows that
\m{
\PRr[\mu_f]{\txt{answer ``$0$''}}{[v_0'], Z_{i_0}=1}
= \PRr[\mu]{X_{i_0,j_0}=a_0}{[v_0], Z_{i_0}=1}
.}
Conditioned on $[v_0]\wedge[Z_{i_0}=1]\wedge[X_{i_0,j_0}=a_0]$, the distribution of $Z$ is the same as conditioned only on $[v_0]\wedge[Z_{i_0}=1]$, as follows from \clmref{cl_invP} and the fact that $Z_{[n]\mset{i_0}}$ is a function of $X_{[n]\mset{i_0}}$.
In other words,
\m[m_Zis1P]{
\forall z\in\01^n:\,
\PRr[\mu]{Z=z}{[v_0], Z_{i_0}=1, X_{i_0,j_0}=a_0}
= \PRr[\mu]{Z=z}{[v_0], Z_{i_0}=1}
.}

In the case of $\Cl P'$ the following trivial analogue of \clmref{cl_invP} holds:
\m{
\Ii[Z\sim\mu_f]{\txt{the answer of $v_0'$}}{Z_{[n]\mset{i_0}}}{[v_0'], Z_{i_0}} = 0
;}
accordingly, 
\m{
\forall z\in\01^n:\,
\PRr[\mu_f]{Z=z}{[v_0'], Z_{i_0}=1, \txt{answer ``$0$''}}
= \PRr[\mu_f]{Z=z}{[v_0'], Z_{i_0}=1}
.}
By \bref{m_alindas} and \bref{m_Zis1P}, this means that $\forall z\in\01^n$:
\m{
\PRr[\mu]{Z=z}{[v_0], Z_{i_0}=1, X_{i_0,j_0}=a_0}
\,=\, \PRr[\mu_f]{Z=z}{[v_0'], Z_{i_0}=1, \txt{answer ``$0$''}}
.}
By \bref{m_req} and the symmetry with respect to the value of $Z_{i_0}$ and the answer,
\mal[m_indst]{
\forall z\in\01^n:&\\
&\PRr[\mu]{Z=z}{[v_0], X_{i_0,j_0}=a_0}
\,=\, \PRr[\mu_f]{Z=z}{[v_0'], \txt{answer ``$0$''}}\\
\txt{and}&\\
&\PRr[\mu]{Z=z}{[v_0], X_{i_0,j_0}=1-a_0}
\,=\, \PRr[\mu_f]{Z=z}{[v_0'], \txt{answer ``$1$''}}
.}

So, for every pair of mutually-corresponding vertices $v_0\in\Cl P$ and $v_0'\in\Cl P'$ that satisfy \bref{m_alindas}, both \bref{m_indst} and \bref{m_reqt} must hold.
This is precisely the statement that we want to use for our inductive argument, as described in the beginning of this part.

Formally, the full argument goes like that:
Let $v_{root}\in\Cl P$ and $v_{root}'\in\Cl P'$ be the roots and assume that the initial distribution of $Z$ is the same -- that is, \bref{m_alindas} is satisfied at the roots.
Let $v_1\in\Cl P$ and $v_1'\in\Cl P'$ be mutually-corresponding sons of the roots; by the above statement, both \bref{m_indst} and \bref{m_reqt} hold at $v_{root}$ and $v_{root}'$.
Note that \bref{m_indst} with respect to the roots implies that \bref{m_alindas} is satisfied with respect to $v_1$ and $v_1'$.
Continuing inductively, we conclude that both \bref{m_indst} and \bref{m_reqt} hold (unconditionally) for all mutually-corresponding non-leaves of $\Cl P$ and $\Cl P'$.

Thus we have shown the following:

\lem[l_P']{Let $g:\01^m\to\set{0,1,*}$.
Let $\Cl P$ be a deterministic protocol that queries bits of $X=(X_1\dc X_n)\in\01^{n\tm m}$.
Let $\mu_g$ be a distribution over $\01^m$, supported on legal input values for $g$, and $\mu=\mu_f\circ\mu_g$ be a distribution over $\01^{n\tm m}$, so that $\mu_f$ is the distribution of $(g(X_1\dc g(X_n))$ when $X\sim\mu$.

Then there exists a protocol $\Cl P'$ that queries bits of $Z\in\01^n$, such that an isomorphism $\Cl M$ maps the protocol tree of $\Cl P$ to a polarised tree representing $\Cl P'$.
Moreover, it holds that
\itstart
\item for every vertex $v$ in the tree of $\Cl P$, the probability of reaching it under $X\sim\mu$ is the same as the probability of reaching $\Cl M(v)$ in $\Cl P'$ under $Z\sim\mu_f$;
\item for every vertex $v$ in the tree of $\Cl P$, the distribution of $(g(X_1)\dc g(X_n)$ taken with respect to $X\sim\mu$ and conditioned upon reaching $v$ is the same as the distribution of $(Z_1\dc Z_n)$ taken with respect to $Z\sim\mu_f$ and conditioned upon reaching $\Cl M(v)$ in $\Cl P'$.
\itend
}

The protocol $\Cl P'$ is, in the first place, a protocol for computing $f(Z)$.
From the above lemma it follows that for every leaf $l$ of $\Cl P$
\m{
\PRr[Z\sim\mu_f]{f(Z)=1}{\Cl M(l)} = \PRr[X\sim\mu]{f\circ g^n(X)=1}{l}
.}
Therefore, if we label every $\Cl M(l)$ by the same answer as appears on $l$, we get a protocol that computes $f$ over $\mu_f$ as accurately as $\Cl P$ computes $f\circ g^n$ over $\mu$.

Now let us assume for a moment that $\mu_f$ is supported on the whole $\01^n$ and revisit \lemref{l_P'}.
According to its statement, if $X\sim\mu$ and $Z\sim\mu_f$, then for every leaf $l$ of $\Cl P$ the distribution of $(g(X_1)\dc g(X_n)$ conditioned upon reaching $l$ is the same as the distribution of $Z$ conditioned upon reaching $\Cl M(l)$ in $\Cl P'$; moreover, the probabilities of reaching $l$ and of reaching $\Cl M(l)$ are the same.
Therefore, for every $z_0\in\01^n$:
\mal{
\PR[X\sim z_0\circ\mu_g]{l}
&=\PRr[\mu_f\circ\mu_g]{l}{Z=z_0}\\
&=\PRr[\mu_f\circ\mu_g]{Z=z_0}{l}
\tm\fr{\PR[\mu_f\circ\mu_g]l}{\PR[\mu_f\circ\mu_g]{Z=z_0}}\\
&=\PRr[Z\sim\mu_f]{Z=z_0}{\Cl M(l)}\tm\fr{\PR[\mu_f]{\Cl M(l)}}{\PR[\mu_f]{Z=z_0}}
=\PRr{\Cl M(l)}{Z=z_0}
}
(note that the rightmost probability only depends on the ``internal randomness'' of $\Cl P'$, and not on the distribution of $Z$).

In other words, the distribution of the leaf that $\Cl P'(Z)$ reaches when $Z=z_0$ is the same as the distribution of $\Cl M(L)$, where $L$ is the leaf reached by $\Cl P(X)$ when $X\sim z_0\circ\mu_g$.\fn
{
Recall that $\Cl P'$ is randomised, and so the leaf reached by the protocol is not necessarily determined by the input value.
Note also that $z_0\circ\mu_g$ is the distribution of $X\sim\mu_f\circ\mu_g$, conditioned upon $Z=(g(X_1)\dc g(X_n))=z_0$.
}
Accordingly, the protocols $\Cl P$ and $\Cl P'$ ``perform identically'' (in the sense of \lemref{l_P'}) even if conditioned upon the value of $Z$, and therefore also with respect to \e{any} distribution $Z\sim\nu$ (corresponding to $X\sim\nu\circ\mu_g$).
In other words, the construction of $\Cl P'$ can be done independently of $\mu_f$:\ it is enough to know the protocol $\Cl P$, the function $g$ and the distribution $\mu_g$.

To conclude:
\crl[crl_P']{Let $f\sbseq\01^n\times\Xi$ and $g:\01^m\to\set{0,1,*}$.
Let $\Cl P$ be a deterministic protocol that queries bits of $X=(X_1\dc X_n)\in\01^{n\tm m}$.
Let $\mu_g$ be a distribution over $\01^m$, supported on legal input values for $g$.

Then there exists a protocol $\Cl P'$ that queries bits of $Z\in\01^n$, such that an isomorphism $\Cl M$ maps the protocol tree of $\Cl P$ to a polarised tree representing $\Cl P'$.
Moreover, for every distribution $\nu$ over $\01^n$ it holds that
\itstart
\item the error of $\Cl P'$ in computing $f(Z)$ when $Z\sim\nu$ is the same as that of $\Cl P$ in computing $f\circ g^n(X)$ when $X\sim\nu\circ\mu_g$;
\item for every $z_0$ in the support of $\nu$, the distribution of $\Cl M(L)$, where $L$ is the leaf reached by $\Cl P$ conditioned upon $(g(X_1)\dc g(X_n))=z_0$ is the same as the distribution of the leaf reached by $\Cl P'$ conditioned upon $Z=z_0$;
\item for every vertex $v$ in the tree of $\Cl P$, the probability of reaching it under $X\sim\nu\circ\mu_g$ is the same as the probability of reaching $\Cl M(v)$ in $\Cl P'$ under $Z\sim\nu$;
\item for every vertex $v$ in the tree of $\Cl P$, the distribution of $(g(X_1)\dc g(X_n))$ taken with respect to $X\sim\nu\circ\mu_g$ and conditioned upon reaching $v$ is the same as the distribution of $Z$ taken with respect to $Z\sim\nu$ and conditioned upon reaching $\Cl M(v)$ by $\Cl P'$.
\itend
}

\sect[s_compl]{Comparing the complexities of the protocols}

In \sref{s_prot} we were given a query protocol $\Cl P$ for computing $f\circ g^n$ and used it to construct a protocol $\Cl P'$ for computing $f$.
Now let us analyse the properties of $\Cl P'$ (as summarised in \crlref{crl_P'}) in order to argue that it has low query complexity if that of $\Cl P$ was low.

Let $\mu_f$ be such input distribution for $f$ that $\R[\mu_f,\eps](f)\in\asOm{\R(f)}$ for some fixed $\eps>0$.
We will assume that $\mu_f$ is \e{non-fixing} in the following sense:\ for any $i_0\in[n]$ and $s_0\in\01^{n-1}$, the distribution of $Z_{i_0}$ when $Z\sim\mu_f$ and $Z|_{[n]\smin\set{i_0}}=s_0$ has positive entropy.\fn
{
If $\mu_f$ is fixing, we let $\eps'\deq\dr\eps2$, $\mu_f'\deq\fr{\mu_f+\eps'\tm\U[\01^n]}{1+\eps'}$ and take $(\mu_f',\eps')$ instead of $(\mu_f,\eps)$:\ the resulting $\mu_f'$ is non-fixing and $\R[\mu_f',\eps'](f)\ge\R[\mu_f,\eps](f)\in\asOm{\R(f)}$.
}
Let $\mu_g$ be input distribution for $g$, supported on legal input values (to be chosen in \sref{sss_mug}; it will be hard in a certain ``error-independent'' sense).

We apply \crlref{crl_P'} with respect to this $\mu_g$ and the given protocol $\Cl P$, letting $\Cl P'$ and $\Cl M$ be as guaranteed by the statement.

\ssect[ss_pred]{Protocols as $Z_i$-predictors}

Informally, in our analysis we will look at the ``knowledge'' of a given leaf of a protocol tree about $Z_{i_0}$.
To formalise this, we consider the behaviour of a protocol with respect to the uniform distribution of $Z$, which corresponds to $X\sim\mu_g^{\U}\deq\U[\01^n]\circ\mu_g$ in the case of $\Cl P$.

Let $\Cl T$ denote the protocol tree of $\Cl P$.
For every leaf $l\in\Cl T$, let $\lambda_{i_0}(l)\in\01$ be a most likely value of $Z_{i_0}=g(X_{i_0})$ when $X\sim\mu_g^{\U}$, conditioned on reaching $l$.\fn
{
Recall that $\mu_g$ is supported only on legal input values for $g$, so $\PRr{g(X_{i_0})=0}{l}+\PRr{g(X_{i_0})=1}{l}=1$ always.
}
Let
\m{
\delta_{i_0}(l)\deq\PRr[X\sim\mu_g^{\U}]{g(X_{i_0})=\lambda_{i_0}(l)}{l}-\fr12
\tb
\lf\{\in\lf[0,\fr12\rt]\rt\}
}
and
\m{
\delta_{i_0}(\Cl T)\deq\E[L]{\delta_{i_0}(L)}
,}
where $L$ is distributed as the leaf of $\Cl T$ reached by $\Cl P$ when $X\sim\mu_f\circ\mu_g$ (note the ``mixture of distributions'':\ $\lambda_{i_0}(l)$ and $\delta_{i_0}(l)$ are defined relative to $X\sim\mu_g^{\U}$, but $L$ in the definition of $\delta_{i_0}(\Cl T)$ is sampled with respect to $X\sim\mu_f\circ\mu_g$).

Let $\Cl T'$ denote the protocol tree of $\Cl P'$.
For every leaf $l'\in\Cl T'$, let $\lambda_{i_0}(l')\in\01$ be a most likely value of $Z_{i_0}$ when $Z\sim\U[\01^n]$, conditioned on reaching $l'$.
Let
\m{
\delta_{i_0}(l')\deq\PRr[{Z\sim\U[\01^n]}]{Z_{i_0}=\lambda_{i_0}(l')}{l'}-\fr12
\tb
\lf\{\in\lf[0,\fr12\rt]\rt\}
}
and
\m{
\delta_{i_0}(\Cl T')\deq\E[L']{\delta_{i_0}(L')}
,}
where $L'$ is the distributed as the leaf of $\Cl T'$ reached by $\Cl P'$ when $Z\sim\mu_f$ (note the ``mixture of distributions'':\ $\lambda_{i_0}(l)$ and $\delta_{i_0}(l)$ are defined relative to $Z\sim\U[\01^n]$, but $L'$ in the definition of $\delta_{i_0}(\Cl T')$ is sampled with respect to $Z\sim\mu_f$).

Note several important symmetries in the above definitions with respect to the isomorphism $\Cl M$.
First, for every leaf $l$ of $\Cl P$
\m{
\lambda_{i_0}(l) = \lambda_{i_0}(\Cl M(l))
\txt{\tb and\tb}
\delta_{i_0}(l) = \delta_{i_0}(\Cl M(l))
.}
Second, the distribution of $\Cl M(L)$, where $L$ is the leaf of $\Cl T$ reached by $\Cl P$ when $X\sim\mu_f\circ\mu_g$ is the same as the distribution of the leaf of $\Cl T'$ reached by $\Cl P'$ when $Z\sim\mu_f$.
Accordingly,
\m[m_TT]{
\delta_{i_0}(\Cl T) = \delta_{i_0}(\Cl T')
.}
These symmetries hold due to the fact that the construction of \crlref{crl_P'} only depends on $\mu_g$ -- in particular, its guarantees hold both in the case of $X\sim\mu_g^{\U}$ and in the case of $X\sim\mu_f\circ\mu_g$.

\sssect{The case of $\Cl P'$}

First we take a closer look at $\Cl T'$, the tree of $\Cl P'$.
It is polarised, so we ``statically'' define $w_{i_0}(l')\in\01$ such that conditioned on reaching the leaf $l'\in\Cl T'$, if $w_{i_0}\neq*$, then $w_{i_0}=w_{i_0}(l')$.
Let
\m{
q_{i_0}(l')
\deq\PRr[{Z\sim\U[\01^n]}]{w_{i_0}\neq*}{l'}
=\PRr[{Z\sim\U[\01^n]}]{\txt{$Z_{i_0}$ has been queried by $\Cl P'(Z)$}}{l'}
.~\footnotemark}
\footnotetext{
Note that if $q_{i_0}(l')=0$, then $w_{i_0}(l')\in\01$ can be defined arbitrarily -- this is similar to the situation when $\delta_{i_0}(l)=0$ in the case of $l\in\Cl P$.
}

Obviously, under $Z\sim\U[\01^n]$ the best guess for the value of $Z_{i_0}$ conditioned on reaching $l'$ would be $w_{i_0}(l')$; therefore,
\m{
\lambda_{i_0}(l')=w_{i_0}(l')
.}
The bits of $Z$ are both unbiased and mutually independent under $\U[\01^n]$; accordingly, if protocol $\Cl P'$ ``knows something'' about $Z_{i_0}$, then $w_{i_0}\neq*$ and the protocol knows that value with certainty:
\m{
\delta_{i_0}(l')
= \fr12\tm(1-q_{i_0}(l')) + q_{i_0}(l') - \fr12
=\fr{q_{i_0}(l')}2
}
and
\m[m_deio]{
\delta_{i_0}(\Cl T')
=\E[L']{\delta_{i_0}(L')}
=\fr12\tm\E[L']{q_{i_0}(L')}
,}
where $L'$ is distributed as the leaf of $\Cl T'$ reached by $\Cl P'$ when $Z\sim\mu_f$.

Next we want to use $\delta_{i_0}(\Cl T')$ as an upper bound on the number of queries made by $\Cl P'(Z)$ to $Z_{i_0}$ under $Z\sim\mu_f$.
The main obstacle here is the fact that $q_{i_0}(l')$ is defined with respect to $Z\sim\U[\01^n]$.\fn
{
Note that the value of $\PRr[Z\sim\nu]{\txt{$Z_{i_0}$ has been queried by $\Cl P'(Z)$}}{l'}$ is, in general, not ``\f\nu-independent'' -- in spite of the fact that the parameters of the generic nodes in $\Cl T'$ are \f\nu-independent, and therefore known.
}

\clm[cl_que-zi]{For every $z\in\01^n$ such that $\Cl P'(z)$ reaches $l'$ with positive probability, it holds that
\m{
\PRr[{Z\sim\U[\01^n]}]
{\txt{$Z_{i_0}$ is queried by $\Cl P'(Z)$}}{[l'],Z_{i_0}=z_{i_0}}
=\PRr{\txt{$Z_{i_0}$ is queried by $\Cl P'(z)$}}{l'}
.}
}

That is, the probability that $Z_{i_0}$ is queried doesn't depend on $Z_{[n]\mset{i_0}}$.
Note that the right-hand side of the above equality only depends on the ``internal randomness'' of $\Cl P'$ (and not on the distribution of $Z$).

\prfstart[\clmref{cl_que-zi}]
Note that 
\m{
\PRr[{\U[\01^n]}]
{\txt{$Z_{i_0}$ is queried by $\Cl P'(Z)$}}{[l'],Z_{i_0}=z_{i_0}}
=\E[\mac{Z'\sim\U[\01^n]\\Z_{i_0}'=z_{i_0}}]
{\PRr{\txt{$Z_{i_0}$ is queried by $\Cl P'(Z')$}}{l'}}
.}
We claim that for every $z'\in\01^n$, the value of
$\PRr{\txt{$Z_{i_0}$ is queried by $\Cl P'(z')$}}{l'}$
is a function of $z_{i_0}'$ -- in particular, this means that the expectation on the right-hand side of the above equality is over a constant value equal to $\PRr{\txt{$Z_{i_0}$ is queried by $\Cl P'(z)$}}{l'}$ (quod erat demonstrandum).

Let $v_1\dc v_t$ be the \f Z-nodes on the path from the root of $\Cl T'$ to $l'$ that may query $Z_{i_0}$, listed in order of appearance.
For $j\in[t]$ let $a_j\in\01$ be the ``answer'' of $v_j$ on the path to $l'$ and let $\fb e_j$ denote the event that $Z_{i_0}$ is queried by $\Cl P'(z')$ in $v_j$.
Let $v_j$ be a \f Z-node, parametrised by $(i_0,\alpha,\beta)$.
Note that conditional on reaching $v_j$, the events $[\wedge_{k=1}^{j-1}(\lnot \fb e_k)]$ and $[w_{i_0}=*]$ coincide.

If $a_j\neq z_{i_0}'$, then $\PR{\fb e_j}=0$.

If $a_j=z_{i_0}'=0$, then
\mal{
&\PRr{\txt{$a_j$ is answered, $\fb e_j$}}
{\txt{$v_j$ is reached, $\wedge_{k=1}^{j-1}(\lnot\fb e_k)$}}
=\alpha,\\
&\PRr{\txt{$a_j$ is answered, $\lnot\fb e_j$}}
{\txt{$v_j$ is reached, $\wedge_{k=1}^{j-1}(\lnot\fb e_k)$}}
=(1-\alpha)\tm(1-\beta)\\
&\tb\Then\tb
\PRr{\fb e_j}
{\txt{$v_j$ is reached, $a_j$ is answered, $\wedge_{k=1}^{j-1}(\lnot\fb e_k)$}}
=\fr\alpha{\alpha+(1-\alpha)\tm(1-\beta)}
.}

If $a_j=z_{i_0}'=1$, then
\mal{
&\PRr{\txt{$a_j$ is answered, $\fb e_j$}}
{\txt{$v_j$ is reached, $\wedge_{k=1}^{j-1}(\lnot\fb e_k)$}}
=\alpha,\\
&\PRr{\txt{$a_j$ is answered, $\lnot\fb e_j$}}
{\txt{$v_j$ is reached, $\wedge_{k=1}^{j-1}(\lnot\fb e_k)$}}
=(1-\alpha)\tm\beta\\
&\tb\Then\tb \PRr{\fb e_j}
{\txt{$v_j$ is reached, $a_j$ is answered, $\wedge_{k=1}^{j-1}(\lnot\fb e_k)$}}
=\fr\alpha{\alpha+(1-\alpha)\tm\beta}
.}

As $a_j$, $\alpha$ and $\beta$ are constants,
\m{
\PRr{\fb e_j}
{\txt{$v_j$ is reached, $a_j$ is answered, $\wedge_{k=1}^{j-1}(\lnot\fb e_k)$}}
}
is a function of $z_{i_0}'$, as well as
\m{
\PRr{\txt{$Z_{i_0}$ is queried by $\Cl P'(z')$}}{l'}
=\sum_{j=1}^t\PRr{\fb e_j}
{\txt{$v_j$ is reached, $a_j$ is answered, $\wedge_{k=1}^{j-1}(\lnot\fb e_k)$}}
,}
and the result follows.
\prfend

Let us decompose
\mal{
q_{i_0}(l')
&=\PRr[{Z\sim\U[\01^n]}]{\txt{$Z_{i_0}$ is queried by $\Cl P'(Z)$}}{l'}\\
&=\E[a\unin\01]
{\PRr[{Z\sim\U[\01^n]}]
{\txt{$Z_{i_0}$ is queried by $\Cl P'(Z)$}}{[l'],Z_{i_0}=a}}
.}
For every $z\in\01^n$ such that $\Cl P'(z)$ reaches $l'$ with positive probability:
\mal{
q_{i_0}(l')
&\ge\fr12\tm\PRr[{Z\sim\U[\01^n]}]
{\txt{$Z_{i_0}$ is queried by $\Cl P'(Z)$}}{[l'],Z_{i_0}=z_{i_0}}\\
&=\fr12\tm\PRr{\txt{$Z_{i_0}$ is queried by $\Cl P'(z)$}}{l'}
,}
where the equality follows from \clmref{cl_que-zi}.
So, for every distribution $\nu$ it holds that
\m{
\PRr[{Z\sim\nu}]{\txt{$Z_{i_0}$ is queried by $\Cl P'(Z)$}}{l'}
=\E[Z'\sim\nu]
{\PRr{\txt{$Z_{i_0}$ is queried by $\Cl P'(Z')$}}{l'}}
\le2\tm q_{i_0}(l')
.}

By \bref{m_deio},
\mal{
\PR[Z\sim\mu_f]{\txt{$Z_{i_0}$ is queried by $\Cl P'(Z)$}}
&=\E[L']{\PRr{\txt{$Z_{i_0}$ is queried}}{L'}}\\
&\le2\tm\E[L']{q_{i_0}(L')}
=4\tm\delta_{i_0}(\Cl T')
,}
where $L'$ is distributed as the leaf of $\Cl T'$ reached by $\Cl P'(Z)$ when $Z\sim\mu_f$.
Then
\m{
\E[Z\sim\mu_f]{\txt{number of queries made by $\Cl P'(Z)$}}
\le4\tm\sum_{i=1}^n\delta_i(\Cl T')
.}
So,
\m[m_delT']{
\sum_{i=1}^n\delta_i(\Cl T')
\ge\fr14\tm\R[\mu_f,\eps](f)
\in\asOm{\R(f)}
.}

\sssect{The case of $\Cl P$}

From \bref{m_delT'} and \bref{m_TT} we have:
\m[m_delT]{
\sum_{i=1}^n\delta_i(\Cl T) \in\asOm{\R(f)}
,}
where $\Cl T$ is the tree of the given protocol $\Cl P(X)$.
Let us analyse $\Cl P$, trying to obtain a lower bound on $\sum\delta_i(\Cl T)$.

For $x\in\01^{n\tm m}$, denote by $l(x)$ the leaf of $\Cl T$ that is reached by $\Cl P(x)$.\fn
{
Note that $l(\dt)$ is well-defined, as $\Cl P$ is deterministic.
}
Let $\mu=\mu_f\circ\mu_g$, then
\m{
\delta_{i_0}(\Cl T)
=\E[X\sim\mu]{\delta_{i_0}(l(X))}
=\E[X'\sim\mu]
{\Ee[X\sim\mu]{\delta_{i_0}(l(X))}
{X_{[n]\mset{i_0}}=X_{[n]\mset{i_0}}'}}
}
(note that $X_{[n]\mset{i_0}}$ contains $(n-1)\tm m$ bits).
Let
\m{
\delta_{i_0}^{(x)}(\Cl T)
\deq\Ee[X\sim\mu]{\delta_{i_0}(l(X))}
{X_{[n]\mset{i_0}}=x_{[n]\mset{i_0}}}
,}
then
\m[m_delTE]{
\delta_{i_0}(\Cl T) = \E[X'\sim\mu]{\delta_{i_0}^{(X')}(\Cl T)}
.}

Let $d_{\mu}(\Cl T)$ denote the expected number of oracle queries that $\Cl P(X)$ makes when $X\sim\mu$ (this is the ``expected depth'' of $\Cl T$).
Let $d_{\mu}^{(i_0)}(\Cl T)$ denote the expected (total) number of queries to bits of $X_{i_0}$ by $\Cl T(X)$ when $X\sim\mu$, and let $d_{\mu}^{(i_0,x)}(\Cl T)$ denote the same expectation, conditioned upon $[X_{[n]\mset{i_0}}=x_{[n]\mset{i_0}}]$.

Obviously,
\m{
d_{\mu}(\Cl T) = \sum_{i=1}^nd_{\mu}^{(i)}(\Cl T)
\txt{\tb and\tb}
d_{\mu}^{(i)}(\Cl T) = \E[X'\sim\mu]{d_{\mu}^{(i,X')}(\Cl T)}
.}
Since we can assume that $d_{\mu}(\Cl T)\in\asO{\R(f\circ g^n)}$,
\m[m_dT]{
\sum_{i=1}^n\E[X'\sim\mu]{d_{\mu}^{(i,X')}(\Cl T)} \in \asO{\R(f\circ g^n)}
.}

\paragraph{Restricting and trimming $\Cl P$}

Now we only miss an upper bound on $\delta_{i_0}^{(x)}(\Cl T)$ in terms of $d_{\mu}^{(i_0,x)}(\Cl T)$ in order to be able to put together \bref{m_delT}, \bref{m_delTE} and \bref{m_dT}.

In this part we construct a ``restriction'' of protocol $\Cl P$, which will compute $g(\dt)$ and whose accuracy and complexity will be closely related to $\delta_{i_0}^{(x)}(\Cl T)$ and $d_{\mu}^{(i_0,x)}(\Cl T)$, respectively.
It will remain to state that if its accuracy is ``noticeable'', then the expected number of queries that it makes cannot be ``negligible'' -- that will be done in \sref{sss_mug} via choosing a suitable distribution $\mu_g$.

Recall the definitions of $\mu_g^0$ and $\mu_g^1$ from \sref{ss_prot}.
Our ultimate $\mu_g$ will be such that
\m[m_mg]{
\mu_g = \fr{\mu_g^0+\mu_g^1}2
}
-- i.e., $g$ will be unbiased with respect to it.
For $i_0\in[n]$ and $x\in\01^{n\tm m}$, denote by $\mu_g^{(i_0,x)}$ the distribution of $X_{i_0}\in\01^m$ when $X\sim\mu=\mu_f\circ\mu_g$, conditioned upon $[X_{[n]\mset{i_0}}=x_{[n]\mset{i_0}}]$.
Note that $\mu_g^{(i_0,x)}$ is a convex combination of $\mu_g^0$ and $\mu_g^1$.

We can easily turn $\Cl T$ into a protocol for computing $g(X_{i_0})$:
Label each leave $l$ of the new protocol by $\lambda_{i_0}(l)$, as defined in the beginning of \sref{ss_pred}.
When $\Cl T$ queries a bit of $X_{i_0}$, the new protocol does the same, and whenever $\Cl T$ queries a bit of $X_{[n]\mset{i_0}}$, the new protocol ``locally'' substitutes the corresponding bit of some fixed $x\in\01^{n\tm m}$ (so, the value of $x$ may affect protocol's behaviour).
Denote this new (deterministic) protocol by $\Cl P^{(i_0,x)}$, let us have a closer look at some of its properties.

Assume that $X_{i_0}\sim\mu_g$.
The protocol $\Cl P^{(i_0,x)}$ queries (only) bits of $X_{i_0}$ and computes $g(X_{i_0})$ with some accuracy.
We would like to use $\delta_{i_0}^{(x)}(\Cl T)$ as a ``measure of accuracy'' of $\Cl P^{(i_0,x)}$ and $d_{\mu}^{(i_0,x)}(\Cl T)$ as its ``measure of complexity''.

Note that the value of $\delta_{i_0}^{(x)}(\Cl T)$ does not directly attest the accuracy of $\Cl P^{(i_0,x)}$ under $X_{i_0}\sim\mu_g$, as $\delta_{i_0}^{(x)}(\Cl T)$ has been defined relative to $\mu_g^{(i_0,x)}$.
Nevertheless,
\mal[P]{
\delta_{i_0}^{(x)}(\Cl T)
&=\E[\mac{X_{i_0}\sim\mu_g^{(i_0,x)}\\X_{[n]\mset{i_0}}=x_{[n]\mset{i_0}}}]
{\delta_{i_0}(l(X))}\\
&=\sum_{a\in\01} \PR[X_{i_0}\sim\mu_g^{(i_0,x)}]{g(X_{i_0})=a}\tm
\Ee[\mac{X_{i_0}\sim\mu_g^{(i_0,x)}\\X_{[n]\mset{i_0}}=x_{[n]\mset{i_0}}}]
{\delta_{i_0}(l(X))}{g(X_{i_0})=a}\\
&\le2\tm\sum_{a\in\01} \fr12\tm
\Ee[\mac{X_{i_0}\sim\mu_g^{(i_0,x)}\\X_{[n]\mset{i_0}}=x_{[n]\mset{i_0}}}]
{\delta_{i_0}(l(X))}{g(X_{i_0})=a}\\
&=2\tm\E[\mac{X_{i_0}\sim\mu_g\\X_{[n]\mset{i_0}}=x_{[n]\mset{i_0}}}]
{\delta_{i_0}(l(X))}
,}
where the equality follows from \bref{m_mg} and the fact that $\delta_{i_0}(l)\ge0$ always.
Therefore,
\m[m_Pidel]{
\PR[X_{i_0}\sim\mu_g]{\Cl P^{(i_0,x)}(X_{i_0})=g(X_{i_0})}
\ge\fr12+\fr{\delta_{i_0}^{(x)}(\Cl T)}2
.}

Relating the query complexity of $\Cl P^{(i_0,x)}$ to the value of $d_{\mu}^{(i_0,x)}(\Cl T)$ is more interesting, as the latter can be much smaller than the expected number of queries made by the protocol under $\mu_g$.\fn
{
E.g., if $\PR{g(X_{i_0})=0}=\dr1m$ under $X_{i_0}\sim\mu_g^{(i_0,x)}$, and $\Cl P^{(i_0,x)}$ makes \asOm m\ queries when $g(X_{i_0})=0$ and \asO1\ queries when $g(X_{i_0})=1$, then $d_{\mu}^{(i_0,x)}(\Cl T)\in\asO1$ but $\Cl P^{(i_0,x)}$ makes \asOm m\ expected queries under $\mu_g$.
}
In order to use both $\delta_{i_0}^{(x)}(\Cl T)$ and $d_{\mu}^{(i_0,x)}(\Cl T)$ as intended, we ``trim'' $\Cl P^{(i_0,x)}$.

Define $\Cl P_{tr}^{(i_0,x)}$ as the protocol obtained from (the tree of) $\Cl P^{(i_0,x)}$, where every vertex $v$ such that $\PRr[\mu_g]{g(X_{i_0})=a}v>\dr34$ for some $a\in\01$ is replaced by a leaf labelled with ``$a$'' (the sub-trees that were under these vertices are dropped).
Like $\Cl P$ and $\Cl P^{(i_0,x)}$, $\Cl P_{tr}^{(i_0,x)}$ is deterministic.

First we analyse the accuracy of $\Cl P_{tr}^{(i_0,x)}$ under $X_{i_0}\sim\mu_g$.
Here the ``worst case'' would be if we have trimmed sub-trees that correctly computed the value of $g(X_{i_0})$, in which case the accuracy in those vertices has reduced from $\dr12+\dr12$ to $\dr12+\dr14$.
From \bref{m_Pidel},
\m[m_Ptidel]{
\PR[X_{i_0}\sim\mu_g]{\Cl P_{tr}^{(i_0,x)}(X_{i_0})=g(X_{i_0})}
\ge\fr12+\fr{\delta_{i_0}^{(x)}(\Cl T)}4
.}

From the definitions, $d_{\mu}^{(i_0,x)}(\Cl T)$ equals the expected number of queries made by $\Cl P^{(i_0,x)}$ when $X_{i_0}\sim\mu_g^{(i_0,x)}$.
Therefore, under the same input distribution the expected number of queries made by $\Cl P_{tr}^{(i_0,x)}$ is at most $d_{\mu}^{(i_0,x)}(\Cl T)$.

Assume without loss of generality that
$\PR[\mu_g^{(i_0,x)}]{g(X_{i_0})=1}\ge\fr12$.
Let $v$ be a non-leaf vertex of $\Cl P_{tr}^{(i_0,x)}$ and $\PR v$ be the probability that it is ``visited'' by $\Cl P_{tr}^{(i_0,x)}$ on input $X_{i_0}$.
From \bref{m_mg} it follows that
\mal[P]{
&\fr{\PR[\mu_g^0]v}{\PR[\mu_g^1]v}
=\fr{\PR[\mu_g]{\txt{$v$ and $g(X_{i_0})=0$}}}
{\PR[\mu_g]{\txt{$v$ and $g(X_{i_0})=1$}}}
=\fr{\PRr[\mu_g]{g(X_{i_0})=0}v}
{\PRr[\mu_g]{g(X_{i_0})=1}v}
\le3;\\
&\PR[\mu_g]v = \fr{\PR[\mu_g^0]v+\PR[\mu_g^1]v}2 \le 2\tm\PR[\mu_g^1]v;\\
&\PR[\mu_g^{(i_0,x)}]v
\ge \PR[\mu_g^{(i_0,x)}]{g(X_{i_0})=1}\tm\PR[\mu_g^1]v
\ge\fr12\tm\PR[\mu_g^1]v;\\
&\PR[\mu_g^{(i_0,x)}]v \ge \fr14\tm\PR[\mu_g]v
.}

The leaves of $\Cl P_{tr}^{(i_0,x)}$ make no queries, so the expected number of queries made by the protocol under $X_{i_0}\sim\mu_g$ is at most $4$ times that number under $\mu_g^{(i_0,x)}$.
In other words,
\m[m_Ptidi]{
\E[X_{i_0}\sim\mu_g]{\txt{number of queries made by $\Cl P_{tr}^{(i_0,x)}(X_{i_0})$}}
\le4\tm d_{\mu}^{(i_0,x)}(\Cl T)
.}

\sssect[sss_mug]{Choosing a suitable $\mu_g$}

We have seen so far that under our assumptions there existed a deterministic protocol that made \asO{d_{\mu}^{(i_0,x)}(\Cl T)} expected queries and computed $g(X_{i_0})$ with accuracy $\dr12+\asOm{\delta_{i_0}^{(x)}(\Cl T)}$ under $X_{i_0}\sim\mu_g$.
It remains to choose $\mu_g$ that would make $g(\dt)$ hard to compute with \e{any} non-trivial advantage over randomly guessing the answer.

\lem[l_ghard]{Let $g:\01^m\to\set{0,1,*}$.
There exists a distribution $\mu_g$, such that for any $\delta>0$, any deterministic protocol computing $g(Y)$ with accuracy $\dr12+\delta$ under $Y\sim\mu_g$ makes
$$\asOm{\delta^2\tm\R(g)}$$
expected queries.}

For us the order of quantifiers in this lemma is crucial:\ the claim holds for the same $\mu_g$ with respect to any $\delta$.
In particular, this means that $g$ is balanced perfectly with respect to $\mu_g$.\fn
{
Which, in turn, means that $\mu_g$ is supported only on legal input values for $g$.
}
Accordingly, a non-trivial \e{deterministic} protocol cannot make less than $1$ expected query (otherwise it would never make a query and had accuracy $\dr12$).
Therefore:
\crl[crl_ghard]{Let $g:\01^m\to\set{0,1,*}$.
There is a distribution $\mu_g$, such that for any $\delta>0$, any deterministic protocol computing $g(Y)$ with accuracy $\dr12+\delta$ under $Y\sim\mu_g$ makes
$$\asOm{\delta^2\tm\R(g)}+1$$
expected queries.
In particular, $\PR{g(Y)=0}=\PR{g(Y)=1}=\dr12$.
}

If we choose such $\mu_g$ to be the ``\f g-part'' of the input distribution $X\sim\mu_f\circ\mu_g$, then from \bref{m_Ptidel} and \bref{m_Ptidi}:
\m[m_dd]{
d_{\mu}^{(i_0,x)}(\Cl T) \in \asOm{\l(\delta_{i_0}^{(x)}(\Cl T)\r)^2\tm\R(g)+1}
\sbseq \asOm{\delta_{i_0}^{(x)}(\Cl T)\tm\sq{\R(g)}}
.}

\prfstart[\lemref{l_ghard}]
Let $\alpha$ be such that for every distribution $\nu$, such that $g$ is balanced with respect to it, there exists $\delta_\nu>0$ and a (deterministic) query protocol $\rho_\nu$ that makes at most $\alpha\tm\delta_\nu^2\tm\R(g)$ expected queries and computes $g(Y)$ with accuracy at least $\dr12+\delta_\nu$ when $Y\sim\nu$.
We need to show that $\alpha\in\asOm1$.
Assume $\alpha\le1$.

Let $d_\nu$ denote the expected number of queries that $\rho_\nu$ makes when $Y\sim\nu$.
Obviously, we can assume that $d_\nu\in\asO{\R(g)}$.
If $g$ is balanced with respect to $\nu$, then any protocol computing it with accuracy $\dr12+\delta_\nu$ must make at least $\delta_\nu$ expected queries; accordingly,
\m[m_dmin]{
\alpha\tm\delta_\nu^2\tm\R(g)\ge d_\nu\ge\delta_\nu
\tb\Then\tb
d_\nu,\,\delta_\nu\ge\fr1{\alpha\tm\R(g)} \ge\fr1{\R(g)}
.}

Let $\mu_g'$ be such that $\R[\mu_g',\fr13](g)\in\asOm{\R(g)}$ and assume without loss of generality that $g$ is balanced with respect to $\mu_g'$.
We will use our assumptions to build a (deterministic) protocol tree $\Cl T_g$ for computing $g(Y)$ with high accuracy under $Y\sim\mu_g'$.
Every non-leaf vertex $v\in\Cl T_g$ will correspond to running a ``weak protocol'' that computes $g(Y)$ with accuracy $\dr12+\delta_v$ with respect to certain distribution $\nu_v$ (the outgoing edges will be labelled by the answer returned by that protocol).
Every leaf will be labelled by the answer that $\Cl T_g$ returns upon reaching it.

The tree is constructed inductively, where at every step we handle one vertex -- that is, we decide whether it will corresponds to a leaf in $\Cl T_g$, and if not, then we assign a weak protocol to this vertex.
A \e{non-handled} vertex can only appear as a son of a vertex that has already been handled (and therefore all his predecessors starting from the root have been handled too).

Let $\Cl T_g^{(i)}$ denote the partial protocol tree constructed at step $i$ ($\Cl T_g^{(0)}$ contains only the root $v_{root}$).
Here is the \ord[i] step of our construction:

\itstart
\item[(a)] Let $v\in\Cl T_g^{(i-1)}$ be a closest to the root non-handled vertex.
Let $\nu_v'$ be the distribution of $Y\sim\mu_g'$, conditioned on reaching $v$ by the protocol described by $\Cl T_g^{(i-1)}$ (this is well-defined, as the actions of the \pss[v] predecessors are known).

If the entropy of $g(Y)$ when $Y\sim\nu_v'$ is at most $\dr12$, let $v$ be a leaf in $\Cl T_g^{(i)}$ and label it by the more likely value of $g(Y)$.

Otherwise, let $\nu_v$ be the ``balanced version'' of $\nu_v'$, defined as $\fr{\nu_v^{(0)}+\nu_v^{(1)}}2$ where $\nu_v^{(a)}$ is the distribution of $Y\sim\nu_v'$, conditioned on $[g(Y)=a]$.
Let $\rho_v$ be a protocol that makes $d_v$ expected queries and computes $g(Y)$ with accuracy at least $\dr12+\sq{\fr{d_v}{\alpha\tm\R(g)}}$ when $Y\sim\nu_v$:\ its existence follows from \bref{m_dmin}.
Let the action of $v$ in $\Cl T_g^{(i)}$ be $\rho_v$, and add to $\Cl T_g^{(i)}$ two (non-handled) sons of $v$, corresponding to the possible answers given by $\rho_v$.

\item[(b)] If there is no non-handled vertices in $\Cl T_g^{(i)}$, stop the construction and let $\Cl T_g\deq\Cl T_g^{(i)}$.

\item[(c)] Let $\wt{\Cl T}_g^{(i)}$ be a modification of $\Cl T_g^{(i)}$, where each non-handled vertex becomes a leaf labelled by a most likely value of $g(Y)$ conditioned on reaching that vertex when $Y\sim\mu_g'$.
If
\m[m_stop]{
\PR[Y\sim\mu_g']{\wt{\Cl T}_g^{(i)}(Y)\ne g(Y)}\le\fr13
,}
stop the construction and let $\Cl T_g\deq\wt{\Cl T}_g^{(i)}$.
\itend

\noindent

First we claim that if this construction halts, then
\m[m_Tg-acc]{
\PR[Y\sim\mu_g']{\Cl T_g(Y)\ne g(Y)}\le\fr13
.}
It is obviously the case if condition \bref{m_stop} has been satisfied; if not, then the construction has aborted at \e{(b)}, which means that all the leaves of $\Cl T_g$ have been created at \e{(a)}.
Then the entropy at every leaf of $\Cl T_g$ is at most $\dr12$ and $\PR{\Cl T_g(Y)\ne g(Y)}<\dr19$ (as $\hbin x\le\dr12\To x\nin[\dr19,\dr89]$, where $\hbin\dt$ denotes the binary entropy function).

Let us argue that the construction halts and analyse the query complexity of $\Cl T_g(Y)$ under $Y\sim\mu_g'$.
For a (deterministic) protocol tree $\Cl T$ that queries bits of $Y$, let
\m{
H_g(\Cl T)
\deq \E[Y\sim\mu_g']{\hh[Y'\sim\mu_g']{g(Y')}{Y'\in l_{\Cl T}(Y)}}
,}
where $\h\dt$ denotes the entropy and $l_{\Cl T}(y)$ is the set of all $y'\in\01^m$ that reach the same leaf of $\Cl T$ as $y$ does.
For $i\ge1$, let $v_i$ be the vertex handled at step $i$ and ``$[v_i]$'' denote the event $[v_i$ is reached by $\wt{\Cl T}_g^{(i)}(Y)]$.
Assume that $v_i$ is not a leaf in $\Cl T_g$, then~\fn
{
Let $\wt{\Cl T}_g^{(0)}$ consist of a single root-leaf vertex $v_{root}$ labelled by ``$1$''.
}
\mal[m_Ti-diff]{
H_g(\wt{\Cl T}_g^{(i-1)}) - H_g(\wt{\Cl T}_g^{(i)})
&=\PR[\mu_g']{v_i}
\tm \l( \hh[\mu_g']{g(Y)}{[v_i]} - \hh[\mu_g']{g(Y)}{[v_i],\rho_{v_i}(Y)} \r)\\
&=\PR[\mu_g']{v_i}
\tm \underbrace
{\l( \h[\nu_{v_i}']{g(Y)} - \hh[\nu_{v_i}']{g(Y)}{\rho_{v_i}(Y)} \r)}%
_{(*)}
.}

Let us estimate $(*)$.
Let $\dr12+\delta_{v_i}$ be the accuracy of $\rho_{v_i}(Y)$ in computing $g(Y)$ over $\nu_{v_i}$.
Denote $\alpha_a=\PR[\nu_{v_i}^{(a)}]{\rho_{v_i}(Y)=1}$ for $a\in\01$, then
\m{
\fr{(1-\alpha_0)+\alpha_1}2=\fr12+\delta_{v_i}
\tb\Then\tb
\alpha_1-\alpha_0 = 2\tm\delta_{v_i}
.}
Note that
\m{
\h[\nu_{v_i}']{g(Y)} = \hbin{\E[A]{\alpha_A}}
\txt{\tb and\tb}
\hh[\nu_{v_i}']{g(Y)}{\rho_{v_i}(Y)}
= \E[A]{\hbin{\alpha_A}}
,}
where $A$ is a Boolean random variable distributed like $g(Y)$ under $Y\sim\nu_{v_i}'$.

Let us use H\"older's simple yet useful ``defect estimation'' for Jensen's inequality, as given in~\cite{B12_The_Var}:
\nfct[f_Hol]{H\"older's estimation}{If $f:[a,b]\to\RR$ is twice continuously differentiable and $X$ is a (discrete) random variable taking values on $[a,b]$, then
\m{
\E{f(X)} - f\l(\E X\r)
= \fr{f''(x_0)}2\tm \l(\E{X^2}-\l(\E X\r)^2\r)
\tb\lf\{=\fr{f''(x_0)}2\tm\Var X\rt\}
}
for some $x_0\in[a,b]$.
}
Applying it with $\hbin\dt$ and $\alpha_A$, we get:\fn
{
Formally speaking, $\hbin['']\dt$ is continuous only on $(0,1)$; if $\alpha_0=0$ or $\alpha_1=1$, this violates the condition of H\"older's estimation, as stated above.
However, the requirements of \fctref{f_Hol} can, obviously, be relaxed by letting $f$ be twice continuously differentiable on $(a,b)$ only and continuous on $[a,b]$.
}
\m{
\h[\nu_{v_i}']{g(Y)} - \hh[\nu_{v_i}']{g(Y)}{\rho_{v_i}(Y)}
\ge \fr{\Inff{x\in(0,1)}{-\hbin['']x}}2
\tm \Var{\alpha_A}
> 2\tm \Var{\alpha_A}
.}
As $v_i$ is not a leaf in $\Cl T_g$,
\m{
\h[\nu_{v_i}']{g(Y)}>\fr12
\tb\Then\tb
\fr1{10}<\PR[\nu_{v_i}']{g(Y)=1}<\fr9{10}
}
and
\m{
\Var{\alpha_A}
=(\alpha_0-\alpha_1)^2 \tm(1-\PR[\nu_{v_i}']{g(Y)=1}) \tm\PR[\nu_{v_i}']{g(Y)=1}
>\fr{\delta_{v_i}^2}3
.}
So, \bref{m_Ti-diff} leads to
\m[m_Ti-prog]{
H_g(\wt{\Cl T}_g^{(i-1)}) - H_g(\wt{\Cl T}_g^{(i)})
> \fr{\PR[\mu_g']{v_i} \tm\delta_{v_i}^2}2
\ge \fr{\PR[\mu_g']{v_i}\tm d_{v_i}}{2\tm\alpha\tm\R(g)}
\ge \fr{\PR[\mu_g']{v_i}}{2\tm(\R(g))^2}
,}
where the last two inequalities follow from \bref{m_dmin}.

Next we apply \bref{m_Ti-diff} to see that our construction of $\Cl T_g$ always halts.
Note that at \e{(a)} we always choose a non-handled vertex closest to the root, so at step $i$ there are at most two ``layers'' of $\Cl T_g^{(i)}$ that contain non-handled vertices.
For $k\in\NN$, define the \ord[k] \e{stage} of construction as the collection of all steps that handle a vertex at depth $k$ (observe that the steps of a stage always form an uninterrupted sequence).

Let $i_0$ and $i_1$ be the first and the last steps of stage $k'$, assume that $k'$ was not the last stage of the construction (recall that now we are proving halting).
Let $V_1\sbseq\Cl T_g^{(i_1)}$ be the set of vertices (at depth $k'$) where our construction has assigned a protocol during one of the steps of stage $k'$, and let $L_1$ be the leaves of $\Cl T_g^{(i_1)}$ at depth $k'$ (i.e., these are vertices with conditional entropy of $g(Y)$ at most $\dr12$).
Observe that $V_1\cupdot L_1\sbseq\Cl T_g^{(i_1)}$ is the set of vertices at depth $k'$.
Then
\mal[m_ki-diff]{
H_g(\wt{\Cl T}_g^{(i_0-1)}) - H_g(\wt{\Cl T}_g^{(i_1)})
&=\sum_{v\in V_1}\PR[\mu_g']{v}
\tm \l( \h[\nu_{v}']{g(Y)} - \hh[\nu_{v}']{g(Y)}{\rho_{v_i}(Y)} \r)\\
&\ge \fr{\PR[\mu_g']{\txt{computation of $\Cl T_g(Y)$ goes through $V_1$}}}
{2\tm(\R(g))^2}
,}
where the inequality is \bref{m_Ti-prog}.

On the other hand, from the assumption that $k'$ was not the last stage of the construction it follows that
\m{
\PR[Y\sim\mu_g']{\wt{\Cl T}_g^{(i_1)}(Y)\ne g(Y)}>\fr13
.}
Since for every $l\in L_1$ it holds that $\PRr{\wt{\Cl T}_g^{(i_1)}(Y)\ne g(Y)}l<\dr19$ (as $\hbin x\le\dr12\To x\nin[\dr19,\dr89]$),
\mal{
\PR[Y\sim\mu_g']{\wt{\Cl T}_g^{(i_1)}(Y)\ne g(Y)}
&\le \PR[\mu_g']{\txt{computation of $\Cl T_g(Y)$ goes through $V_1$}} \tm\fr12\\
&\tbbb+\PR[\mu_g']{\txt{computation of $\Cl T_g(Y)$ goes through $L_1$}} \tm\fr19\\
&=\fr19 +\PR[\mu_g']{\txt{computation of $\Cl T_g(Y)$ goes through $V_1$}}\tm\fr7{18}
,}
and therefore,
\m{
\PR[\mu_g']{\txt{computation of $\Cl T_g(Y)$ goes through $V_1$}} > \fr47
.}
From \bref{m_ki-diff},
\m{
H_g(\wt{\Cl T}_g^{(i_0-1)}) - H_g(\wt{\Cl T}_g^{(i_1)})
>\fr2{7\tm(\R(g))^2}
}
and our construction halts after finitely-many steps (as every stage is obviously finite).

It remains to analyse the query complexity of $\Cl T_g$ under $\mu_g'$.
Let $V\sbs\Cl T_g$ be the set of non-leaves and $S\sbs\NN$ be the steps of the construction where the elements of $V$ were handled ($|V|=|S|$).
Then
\m{
H_g(\wt{\Cl T})=
H_g(\wt{\Cl T}_g^{(0)})
-\sum_{i\in S} \l(H_g(\wt{\Cl T}_g^{(i-1)}) - H_g(\wt{\Cl T}_g^{(i)})\r)
\le 1- \fr1{2\tm\alpha\tm\R(g)}\tm\sum_{v\in V}\PR[\mu_g']{v}\tm d_{v}
,}
where the inequality is \bref{m_Ti-prog}, and
\m{
\sum_{v\in V}\PR[\mu_g']{v}\tm d_{v} \le 2\tm\alpha\tm\R(g)
.}
The left-hand side of this inequality is the expected number of queries that $\Cl T_g(Y)$ makes when $Y\sim\mu_g'$.
The result follows from \bref{m_Tg-acc} and the assumption that $\R[\mu_g',\fr13](g)\in\asOm{\R(g)}$.
\prfend[\lemref{l_ghard}]

\ssect{Summing up: the complexities}

From \bref{m_delT} and \bref{m_delTE},
\m{
\sum_{i=1}^n \E[X\sim\mu]{\delta_{i}^{(X)}(\Cl T)} \in\asOm{\R(f)}
.}
From \bref{m_dd}, for all $i\in[n]$ and $x\in\01^{n\tm m}$:
\m{
d_{\mu}^{(i,x)}(\Cl T) \in \asOm{\delta_{i}^{(x)}(\Cl T)\tm\sq{\R(g)}}
.}
Accordingly,
\m{
\sum_{i=1}^n \E[X\sim\mu]{d_{\mu}^{(i,X)}(\Cl T)}
\in\asOm{\sq{\R(g)}} \tm \sum_{i=1}^n \E[X\sim\mu]{\delta_{i}^{(X)}(\Cl T)}
\sbseq{\asOm{\R(f)\tm\sq{\R(g)}}}
.}
By \bref{m_dT}, this implies that
\m{
\asOm{\R(f)\tm\sq{\R(g)}} \cap \asO{\R(f\circ g^n)} \neq\emptyset
.}

To conclude:
\theo[t_main]{Let $f\sbseq\01^n\times\Xi$ and $g:\01^m\to\set{0,1,*}$.
Then
\m{
\R(f\circ g^n) \in \asOm{\R(f)\tm\sq{\R(g)}}
.}
}

\sect[s_tight]{Tightness: $\R(f\circ g^n)\in\asO{\R(f)\tm\sq{\R(g)}}$ is possible}

We construct a relation $f_0\sbseq\01^n\times\01^n$ (i.e., $\Xi=\01^n$) and a promise function $g_0:\01^n\to\set{0,1,*}$ (i.e., $m=n$), such that $\R(f_0)\in\asT{\sq n}$, $\R(g_0)\in\asT{n}$ and $\R(f_0\circ g_0^n)\in\asT{n}$.

Let
\m{
f_0(z)\deq\sett a{\sz{a+z}\le\fr n2-\sq n}
}
and
\m{
g_0(x)\deq\thrcase
{0}{if $\sz x\le\dr n2-\sq n$;}
{1}{if $\sz x\ge\dr n2+\sq n$;}
{*}{otherwise.}
}

\clm[c_fl]{$\R(f_0)\in\asOm{\sq n}$.}

\prfstart
Assume that a deterministic protocol of cost $k$ solves $f_0$ with respect to the uniform input distribution with error at most $\dr14$.
Such protocol partitions $\01^n$ into (at most) $2^k$ sub-cubes, each marked by some ``answer'' (an element from $\01^n$).
In particular, more than $2^{n-1}$ points belong to sub-cubes of size at least $2^{n-k-1}$ -- in other words, to sub-cubes of co-dimension at most $k+1$.
As more than half of all points belong to such sub-cubes and the total protocol error is at most $\dr14$, there exists at least one sub-cube of co-dimension at most $k+1$, on which the protocol errs with probability less than $\dr12$.

The symmetry in the definition of $f_0$ allows us to assume without loss of generality that the sub-cube is the set $\tau\deq0^{k+1}\circ\01^{n-k-1}$, where ``$\circ$'' denotes string concatenation.
It is easy to see that the ``answer'' that would minimise the error probability with respect to this sub-cube can be any binary string starting with ``$0^{k+1}$'', so let us assume that the actual label is $0^n$.
Then
\m{
\PRr{\tit{error}}{Z\in\tau}
=\PR[Z'\unin\01^{n-k-1}]{\sz{Z'}\le\fr n2-\sq n}
<\fr12
,}
which implies that $k+1\ge2\sq n$, as a uniformly-random binary string of length more than $n-2\sq n$ would have more than $\dr n2-\sq n$ ``ones'' with probability at least $\dr12$.
\prfend[\clmref{c_fl}]

\clm[c_gl]{$\R(g_0)\in\asOm n$.}

\prfstart
A randomised query protocol of cost $k$ for $g_0$ would trivially imply existence of a randomised communication protocol of cost at most $2k$ for the bipartite problem \e{Gap-Hamming-Distance}:
\m{
GHD(X,Y)\deq\thrcase
{0}{if $\sz{X\xor Y}\le\dr n2-\sq n$;}
{1}{if $\sz{X\xor Y}\ge\dr n2+\sq n$;}
{*}{otherwise,}
}
and it has been demonstrated by Chakrabarti and Regev~\cite{CR11_An_O} that the complexity of this problem is \asOm n.
\prfend[\clmref{c_gl}]

\clm[c_fgu]{$\R[\eps](f_0\circ g_0^n)\in\asO{n\tm\sq{\log(\dr1\eps)}}$.}

\prfstart
Consider the following protocol for computing $f_0\l(g_0(x_1)\dc g_0(x_n)\r)$, where $x_i\in\01^n$:
For every $i\in[n]$, let $a_i=x_i(j_i)$, where $j_i\unin[n]$ -- that is, $a_i$ is a uniformly-random bit of $x_i$.
Then $\sszz{i}{a_i=g_0(x_i)}$ -- the expected number of ``correctly guessed'' \pl[a_i] is at least $\dr n2+\sq n$; intuitively, this means that the probability that $a_1\dc a_n$ is a right answer to $f_0\l(g_0(x_1)\dc g_0(x_n)\r)$ is ``non-trivially high'' -- to ``boost'' this probability, we will use several ``probes'' from every $x_i$ and take their majority vote.

\noindent
\bil{Protocol:}
For an odd integer $t_\eps$ as defined next, independently choose $j_{i,k}\unin[n]$ for $i\in[n]$ and $k\in[t_\eps]$.
Let $a_i\deq maj\l(x_i(j_{i,1})\dc x_i(j_{i,t_\eps})\r)$ and output ``$a_1\dc a_n$''.

To analyse it, we consider for every $i\in[n]$:
\m{
&\PR{a_i=g_0(x_i)}-\PR{a_i\ne g_0(x_i)}\\
&\tb\ge\sum_{i=0}^{\fr{t_\eps-1}2}
\chs{t_\eps}i\tm
\l(
\l(\fr12-\fr1{\sq n}\r)^i \l(\fr12+\fr1{\sq n}\r)^{t_\eps-i}
-\l(\fr12-\fr1{\sq n}\r)^{\fr{t_\eps+1}2+i}
\l(\fr12+\fr1{\sq n}\r)^{\fr{t_\eps-1}2-i}
\r)\\
&\tb=
\l(1-\l(\fr{1-\dr2{\sq n}}{1+\dr2{\sq n}}\r)^{\fr{t_\eps+1}2}\r)
\tm\sum_{i=0}^{\fr{t_\eps-1}2}
\chs{t_\eps}i\tm
\l(\fr12-\fr1{\sq n}\r)^i \l(\fr12+\fr1{\sq n}\r)^{t_\eps-i}
,}
where the equality occurs when $\sz{x_i}-\dr n2=\pm\sq n$.
As
\m{
\sum_{i=0}^{\fr{t_\eps-1}2}
\chs{t_\eps}i\tm
\l(\fr12-\fr1{\sq n}\r)^i \l(\fr12+\fr1{\sq n}\r)^{t_\eps-i}
=\PRr{a_i=g_0(x_i)}{\sz{x_i}-\fr n2=\pm\sq n}
>\fr12
,}
we get
\m{
&\PR{a_i=g_0(x_i)}-\PR{a_i\ne g_0(x_i)}\\
&\tb>\fr12\tm
\l(1-\l(\fr{1-\dr2{\sq n}}{1+\dr2{\sq n}}\r)^{\fr{t_\eps+1}2}\r)
>\fr12\tm\l(1-\l(1-\fr2{\sq n}\r)^{\dr{t_\eps}2}\r)
\ge\Min{\fr{t_\eps}{4\sq n},\fr14}
.}
Our $t_\eps$ will be small enough to guarantee that $\fr{t_\eps}{4\sq n}\le\fr14$, so we can write
\m[m_aig0]{
\PR{a_i=g_0(x_i)} > \fr12 + \fr{t_\eps}{8\sq n}
.}

Now let us estimate the probability that $a_1\dc a_n$ is a wrong answer to $f_0\l(g_0(x_1)\dc g_0(x_n)\r)$:
This occurs only if $\sszz{i}{a_i=g_0(x_i)}<\dr n2+\sq n$, so by the Chernoff bound (in a form given in~\cite{DM05_Con_Bo}),
\m{
\PR{\txt{the protocol errs}}
<\exp\l(-\fr12\tm\l(\fr{t_\eps}8-1\r)^2\r)
,}
so that choosing $t_\eps\in\asT{\sq{\log(\dr1\eps)}}$ would suffice for our needs and the result follows.
\prfend[\clmref{c_fgu}]

From \theoref{t_main} and Claims \ref{c_fl}, \ref{c_gl} and \ref{c_fgu}:
\theo[t_match]{For $f_0$ and $g_0$ as defined above,
\m{
\R(f_0)\in\asT{\sq n},\,
\R(g_0)\in\asT{n}
\txt{~and~}
\R(f_0\circ g_0^n)\in\asT{n}
.}
}

\sect[s_con]{Conclusions}

We have seen that
\f{\R(f\circ g^n) \in \asOm{\R(f)\tm\sq{\R(g)}}}
for every relation $f$ and promise function $g$, and this can be tight.

One may attempt to prove a more general lower bound by allowing $g$ to be a relation as well (although the corresponding definition of the composed problem looks somewhat artificial).

On the other hand, it may be interesting to analyse the tightness of this lower bound in the following two more restricted cases:\itemi{
\item when both $f$ and $g$ are promise functions;
\item when both $f$ and $g$ are total functions.
}
Depending on the answer, addressing this question may require either proving a stronger lower bound on the complexity of the composed problem or finding a tightness-witnessing example that would use more restricted type of computational problems than what we have seen in \sref{s_tight} (or -- somewhat less likely -- both).

\end{document}